\documentclass[prl, onecolumn]{revtex4}
\usepackage{ae, aecompl}
\usepackage{helvet}
\usepackage{epsfig}
\usepackage{dcolumn}
\usepackage{color}
\usepackage[LGR, T1]{fontenc}
\usepackage[latin9]{inputenc}
\usepackage{amssymb}
\usepackage{amsmath}
\usepackage{mwe, tikz}
\usepackage{esint}
\usepackage{longtable}
\usepackage{url}
\usepackage{etoolbox}
\usepackage{multirow}
\usepackage{graphicx} 

\usepackage{epstopdf} 

\newcommand{\beginsupplement}{%
 \setcounter{table}{0}
 \renewcommand{\thetable}{S\arabic{table}}%
 \setcounter{figure}{0}
 \renewcommand{\thefigure}{S\arabic{figure}}%
 }

\renewcommand\footnotemark{}

\makeatletter
\def\blfootnote{\gdef\@thefnmark{}\@footnotetext}
\makeatother

\makeatletter
\renewcommand{\fnum@figure}{\textbf{Figure \thefigure}}
\makeatother

\usepackage{graphicx}
\usepackage{rotating}

\newcommand*{\grabto}[2]{\IfFileExists{#2}{}{\immediate\write18{curl \detokenize{#1 -o #2}}}}

\DeclareFontEncoding{LGR}{}{}
\DeclareTextSymbol{\~}{LGR}{126}

\begin{document}
\bibliographystyle{apsrev}

\title{A peculiar low-luminosity short gamma-ray burst from a double neutron star merger progenitor}
\author{B.-B. Zhang$^{1,2,3}$, 
B. Zhang$^{4,5,6}$,
H. Sun$^{7}$,
W.-H. Lei$^{8}$,
H. Gao$^{9}$, 
Y. Li$^{6}$,
L. Shao$^{10,11}$,
Y. Zhao$^{12}$,
Y.-D. Hu$^{2, 13}$, 
H.-J. L\"u$^{14}$,
X.-F. Wu$^{11, 15,}$,
X.-L. Fan$^{16}$,
G. Wang$^{17,18}$
A. J. Castro-Tirado$^{2,19}$, 
S. Zhang$^{10}$,
B.-Y. Yu$^{10}$,
Y.-Y. Cao$^{10}$,
E.-W. Liang$^{14}$}

\blfootnote{\footnotesize{
$^{1}$School of Astronomy and Space Science, Nanjing University, Nanjing 210093, China; zhang.grb@gmail.com
$^{2}$Instituto de Astrof\'isica de Andaluc\'ia (IAA-CSIC), P.O. Box 03004, E-18080 Granada, Spain
$^{3}$Key Laboratory of Modern Astronomy and Astrophysics (Nanjing University), Ministry of Education,China
$^{4}$Department of Physics and Astronomy, University of Nevada, Las Vegas, NV 89154, USA; zhang@physics.unlv.edu
$^{5}$Department of Astronomy, School of Physics, Peking University, Beijing 100871, China;
$^{6}$Kavli Institute for Astronomy and Astrophysics, Peking University, Beijing 100871;
$^{7}$National Astronomical Observatories, Chinese Academy of Sciences, A20 Datun Road, Beijing 100012, China;
$^{8}$School of Physics, Huazhong University of Science and Technology, Wuhan 430074, China;
$^{9}$Department of Astronomy, Beijing Normal University, Beijing 100875, China;
$^{10}$Department of Space Sciences and Astronomy, Hebei Normal University, Shijiazhuang 050024, China;
$^{11}$Purple Mountain Observatory, Chinese Academy of Sciences, Nanjing 210008, China;
$^{12}$ Department of Astronomy, University of Florida, 211 Bryant Space Science Center, Gainesville, 32611, USA
$^{13}$Universidad de Granada, Facultad de Ciencias Campus Fuentenueva s/n E-18071 Granada, Spain;
$^{14}$Guangxi Key Laboratory for Relativistic Astrophysics, Department of Physics, Guangxi University, Nanning 530004, China;
$^{15}$Joint Center for Particle, Nuclear Physics and Cosmology, Nanjing University-Purple Mountain Observatory, Nanjing 210008, China;
$^{16}$School of Physics and Electronics Information, Hubei University of Education, Wuhan 430205, China;
$^{17}$Gran Sasso Science Institute (INFN), Via Francesco Crispi 7 LAquila, I-67100, Italy 
$^{18}$INFN - Sezione di Pisa Edificio C, Largo Bruno Pontecorvo, 3, Pisa, 56127, Italy 
$^{19}$Departamento de Ingenier\'ia de Sistemas y Autom\'atica, Escuela de Ingenier\'ias, Universidad de M\'alaga, C\/. Dr. Ortiz Ramos s\/n, 29071 M\'alaga, Spain;}}
 \maketitle

\textbf{Double neutron star (DNS) merger events are promosing candidates of short Gamma-ray Burst (sGRB) progenitors as well as high-frequecy gravitational wave (GW) emitters. On August 17, 2017, such a coinciding event was detected by both the LIGO-Virgo gravitational wave detector network as GW170817 and Gamma-Ray Monitor on board NASA's {\it Fermi} Space Telescope as GRB 170817A. Here we show that the fluence and spectral peak energy of this sGRB fall into the lower portion of the distributions of known sGRBs. Its peak isotropic luminosity is abnormally low. The estimated event rate density above this luminosity is at least $190^{+440}_{-160} {\rm Gpc^{-3} \ yr^{-1}}$, which is close to but still below the DNS merger event rate density. This event likely originates from a structured jet viewed from a large viewing angle. There are similar faint soft GRBs in the {\it Fermi} archival data, a small fraction of which might belong to this new population of nearby, low-luminosity sGRBs. }
\\

Short-duration gamma-ray bursts have long been proposed to be produced in systems involving the coalescence of double neutron stars (DNS) \cite{eichler89}, and the observations of sGRB afterglows and host galaxies are consistent with such a conjecture \cite{gehrels05,fong10,berger14}. 
Based on the estimated event rate density derived from previously observed sGRBs at cosmological distances \cite{wanderman15,sun15}, the chance of detecting a sGRB within a small volume for detectable DNS mergers by advanced LIGO is very low \cite{ghirlanda16}. Thus GRB 170817A\cite{gbm_apjl}/GW 170817\cite{prl}, as the first event in
history showing a GRB associated with a gravitation wave signal from a
compact binary merger, provides an unique opportunity to study its event rate, merger product and implications of the GRB physics. 

In this work, we perform a comprehensive analysis on GRB 170817A, mainly focusing on its prompt emission data in $\gamma$-ray energy band. Taking NGC 4993 as its host galaxy\cite{host-galaxy} we find
that its luminosity is abnormally low. We calculate the event rate of such sGRB event rate density and perform a comparison between such rate density and the NS-NS merger event rate density. We also discuss the possible jet geometries, the physical implication of the time delay between GW signal and GRB signal and the possible merger products of the event.

\begin{figure}
\begin{tabular}{c}
\includegraphics[keepaspectratio, clip, width=0.50\textwidth]{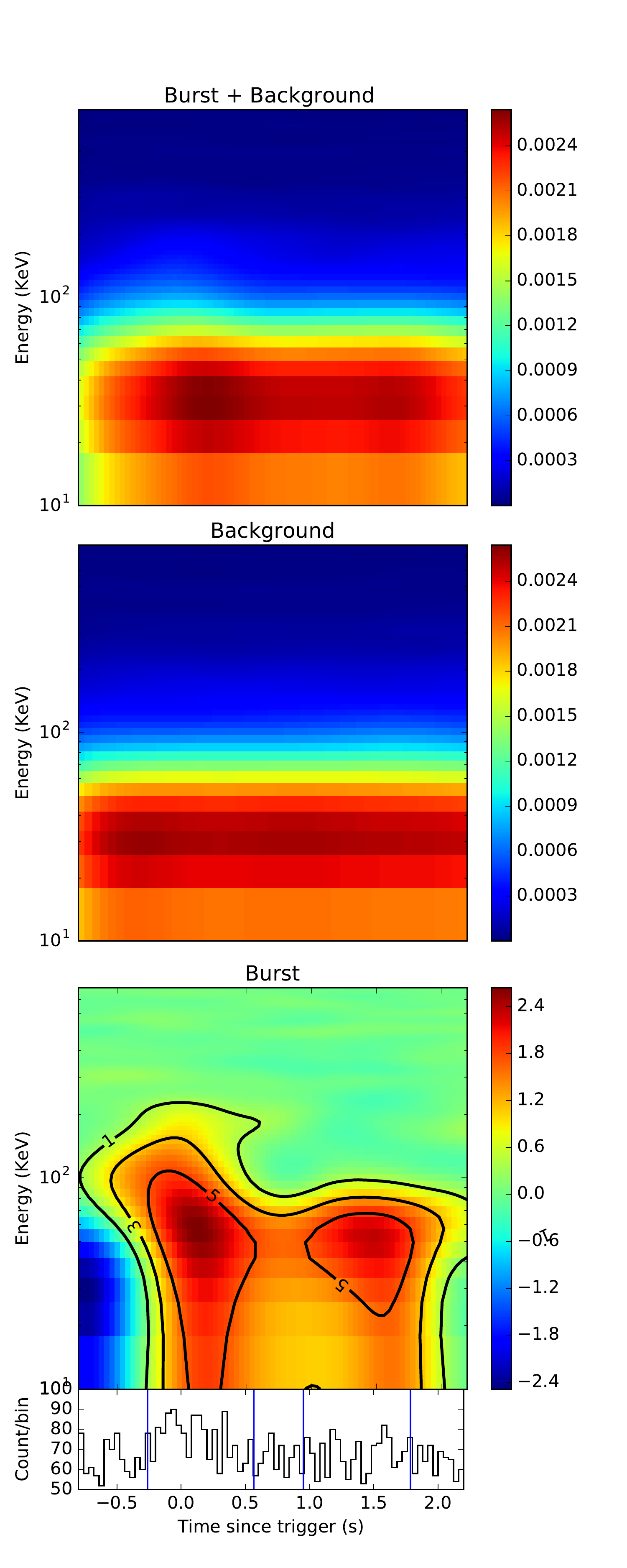} \\
\end{tabular}
\caption{Signal detection from the {\it Fermi \/}GBM Time-Tagged Event (TTE) data of GRB 170817A. {\it a:} the observed count map; {\it b:} the count map in a background region. {\it c:} the background-subtracted count map along with the 15-350 keV light curve. The contour lines represent the levels of signal-to-noise ratio.}
\label{fig:multilc}
\end{figure}

\section{Results}

\noindent\textbf{Light curve structure.}
GRB 170817A ({\it Fermi} Trigger number 170817529) triggered Fermi GBM (8 keV - 40 MeV) \cite{meegan09} 
at $T_0$=12:41:06.474598 UT on 17 August 2017 \cite{gbm_apjl}. We process the public Fermi/GBM data using the procedure as described in \cite{zhang16}. We selected two GBM/NaI detectors, n1 and n2, on board {\it Fermi} that are in good geometric configurations (e.g., angle $<$ 60 deg) with respect to the source position. By extracting the photon events from the Time-Tagged Event (TTE) data detected by these two detectors, we noticed that a sharp peak is present in the light curve between $T_0-0.26$ s and $T_{0}+0.57$ s with a signal-to-noise ratio (S/N) $>$ 5 ({\bf Methods}). Such a signal is clearly identified in the 2-D count map presented in Figure 1. A weaker tail, which is also significant above the background with S/N$>$ 5, appears between $T_0+0.95$ s and $T_0+1.79$ s. The total span of GRB 170817A is about 2.05 s with a 0.38 s-gap consistent with the background.
The burst was also detected in the data of the SPI Anti-Coincidence System (ACS) on-board International Gamma-Ray Astrophysics Laboratory (INTEGRAL)\cite{INTEGRAL}. We download the pre-binned (50-ms bin) SPI-ACS light curve from http://isdc.unige.ch/Soft/ibas/ibas\_acs\_web.cgi, which is derived from 91 independent detectors with different lower energy thresholds (from 60 to 120 keV) and an upper threshold of $\sim$ 10 MeV. 
The multi-channel GBM light curves and the SPI-ACS light curves are presented in {Supplementary Figure 1}. {We performed an extended search for signals before and after the burst using GBM data and no significant emission was found (Supplementary Note 5; Supplementary Figure 9\& 10). }

\noindent\textbf{Spectral analysis.}
We first extract the time integrated spectrum in the first peak region (i.e., from $T_0-0.26$ to $T_0+0.57$ s). We select the NaI detectors n1 \& n2 and BGO detector b0. 
The total number of photon counts is significantly above the background counts in NaI detectors ({Supplementary Figure 11}). {We used a software package developed by the first author, McSpecFit\cite{zhangbb2017natureastro}, to perform spectral fitting. The energy channels at and around the iodine K-edge at 33.17 keV\cite{2012ApJS..199...19G} were excluded to better assess the quality of the fitting of spectral models}. We find that the net spectrum can be successfully fitted by a power law function with an exponential\
high-energy cutoff (hereafter, cutoff power law or CPL model) with the goodness of statistics {PGSTAT= 252.7 and degree of freedom DOF =351} (Supplementary Note 1). The power law index is {-0.61$_{-0.60}^{+0.34}$ and
the cutoff energy, parameterized as $E_p$, is 149.1$_{-24.2}^{+229.4}$ keV.} The corresponding average flux in this time interval is $2.19_{-0.62}^{+3.76} \times 10^{-7} \ {\rm erg \ cm^{-2} \ s^{-1}}$ {between 10 keV and 10 MeV}. {The fluence is $1.81_{-0.51}^{+3.11} \times 10^{-7} \ {\rm erg \ cm^{-2}}$}. {For the second peak between $T_0+0.95$ s and $T_0+1.79$ s, we find that the net spectrum can be preferably fitted by a blackbody model with {$kT=11.3_{-2.4}^{+3.8}$} keV {(PGSTAT/dof=236.4/352)} {although we cannot rule out its non-thermal origin due to the large uncertainty of the lower spectral index when fitted by a cutoff-PL model (Supplementary Table 1)}. The corresponding average flux in this time interval is $5.2_{-2.4}^{+4.7} \times 10^{-8} \ {\rm erg \ cm^{-2} \ s^{-1}}$ {between 10 keV and 10 MeV}. The fluence in the same energy range is $4.33_{-1.99}^{+3.95} \times 10^{-8} {\rm erg \ cm^{-2}}$. Including both peaks, the total fluence is $2.24_{-0.53}^{+3.51} \times 10^{-7} {\rm erg \ cm^{-2}}$, corresponding to an isotropic energy of $4.17_{-0.99}^{+6.54} \times 10^{46}$ erg. 
Using a 50 ms time resolution light curve, the peak luminosity at $T_0\simeq-0.07$ s is derived as $1.6_{-0.4}^{+2.5} \times 10^{47} {\rm erg \ s^{-1}}$.}
The best-fit parameters are presented in Supplementary Table I. The spectral fitting plots as well as the parameter constraints are presented in Supplementary Figures 2-4. {No significant spectral evolution }is observed (Supplementary Figure 5). 

\noindent\textbf{Spectral lag analysis.}
{Using the Cross Correlation Function (CCF) method}, we also calculate the spectral lag of the GRB between (25-50) keV and (50-100) keV, which is $0.03\pm 0.05$ s, consistent with zero. This is consistent with the spectral lag distribution of sGRBs \cite{gehrels06}.

\begin{table}
\caption{\label{tab:sum}Properties of GRB 170817A.}
\label{tab:sum}
\begin{center}
\begin{tabular}{c|c}
\hline
total spanning duration ($s$) & $\sim$2.05\\
spectral peak energy (first peak) $E$$_{\rm p}$ (keV) & 149.1$_{-24.2}^{+229.4}$ \\
total fluence ($erg$ $cm$$^{-2}$) &$2.24_{-0.53}^{+3.51} \times 10^{-7}$ \\
{spectral lag (25-50 $keV$ vs 50-100 $keV$ )} & $0.03\pm 0.05 $ $s$ \\
redshift $z$ & $\sim 0.009$ \\
luminosity distance $D_{\rm L}$ ($Mpc$) & 39.472 \\
total isotropic energy $E_{\rm iso}$ ($erg$) & $4.17_{-0.99}^{+6.54} \times 10^{46} $\\
peak luminosity $L_{\rm iso}$ ($erg$ $s^{-1}$) & $1.6_{-0.4}^{+2.5} \times 10^{47} $ \\
\hline
\end{tabular}
\end{center}
\end{table}

\noindent\textbf{Comparison with other GRBs.}
With the observed and derivative properties summarized in Table \ref{tab:sum}, one can compare GRB 170817A with other sGRBs.
The following samples extracted from the Fermi/GBM catalog \cite{GBM} are considered for comparison: {\it a}. the long GRB sample with $E_p$ measured (1679 GRBs); {\it b}. the short GRB sample (T$_{90}<2$ s) with $E_p$ measured (317 GRBs); and {\it c}. the short GRB sample with S/N < 6 and $E_p$ measured (66 {``faint \& short"} GRBs). The latter is the faint sGRB sample to which GRB 170817A belongs (Supplementary Note 2 \& Supplementary Figure 6).

We first compare the observed properties of GRB 170817A with other GRBs. Figure \ref{fig:fluence_ep} upper panel is the standard $T_{90}-$HR (hardness ratio) plot for GRBs. {One can see that GRB 170817A falls in the boundary between short and long GRB populations. Since evidence has suggested that majority of sGRBs are consistent with the compact star merger origin, GRB 170817A, being associated with GW170817, belongs to the long and soft regime of this population.} Figure \ref{fig:fluence_ep} lower panel 
compares GRB 170817A and other GRBs in the fluence vs. $E_{p}$ diagram. {GRB 170817A seems to lie far away from the majority of the long GRBs}. Based on $\gamma$-ray information only, this burst would be {more likely} regarded as one of those normal (but faint and soft) short GRBs if there were no gravitational wave trigger. Comparing the host galaxy NGC 4993 of GRB 170817A with the host galaxies of other sGRBs \cite{fong10,li16}, we find that NGC 4993 falls into the distribution of sGRB hosts in terms of half-light radius, stellar mass, and afterglow offset from the host galaxy (Supplementary Note 3; Supplementary Figure 7).

\begin{figure}

\begin{tabular}{cc}
\includegraphics[keepaspectratio, clip, width=0.75\textwidth]{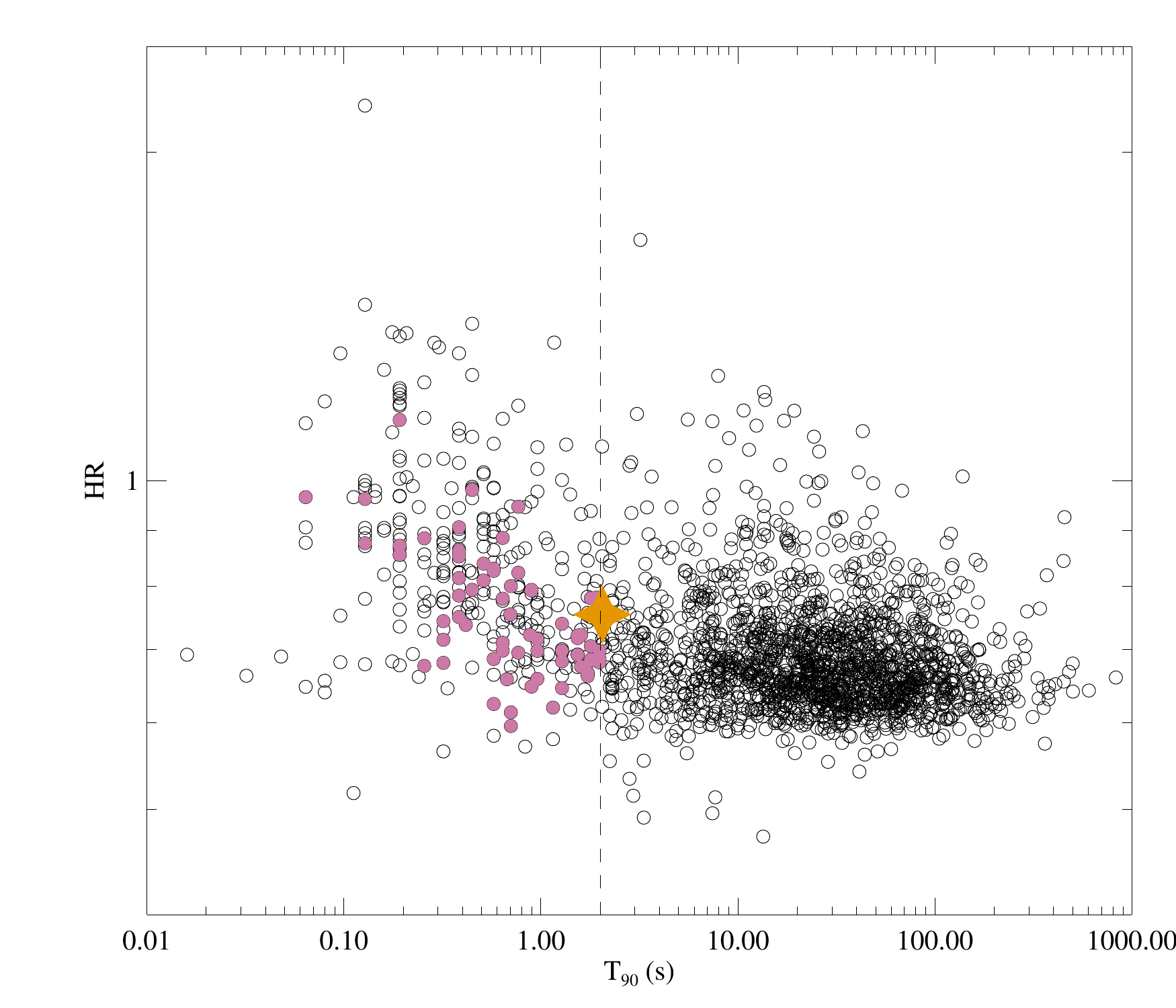} \\
\includegraphics[keepaspectratio, clip, width=0.75\textwidth]{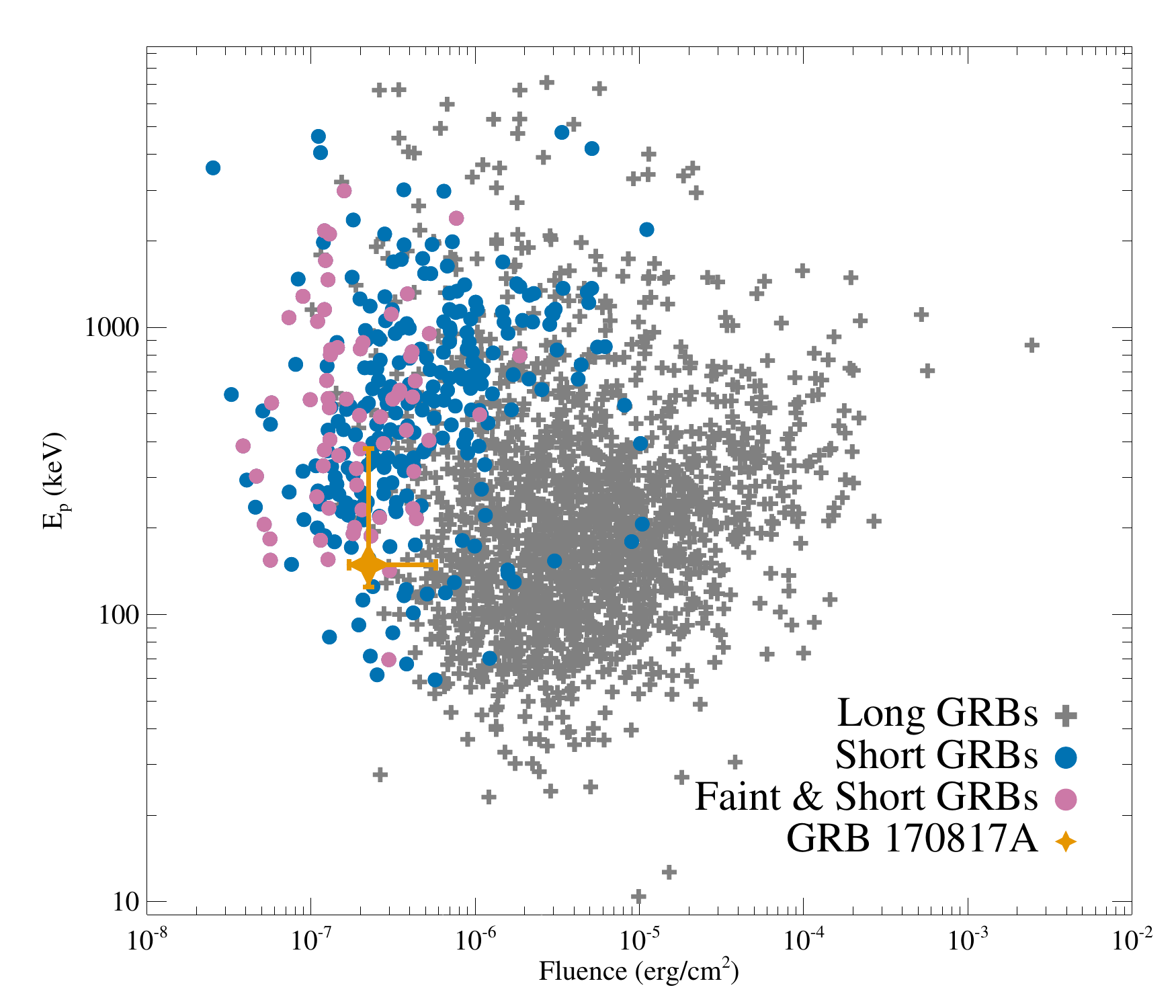} \\
\end{tabular}
\caption{Comparisons between GRB 170817A and other GRBs. {\it a:} A comparison between GRB 170817A and other Fermi long and short GRBs in the $T_{90}-$HR diagram. The hardness ration (HR) is defined as ratio of the observed counts in 50-100 keV band compared to the counts in the 25-50 keV band within the T$_{90}$ region. {\it b:} GRB 170817A in the fluence vs E$_{p}$ diagram against other sGRBs.}
\label{fig:fluence_ep}
\end{figure}

\begin{figure}
\begin{tabular}{c}
\includegraphics[keepaspectratio, clip, width=0.86\textwidth]{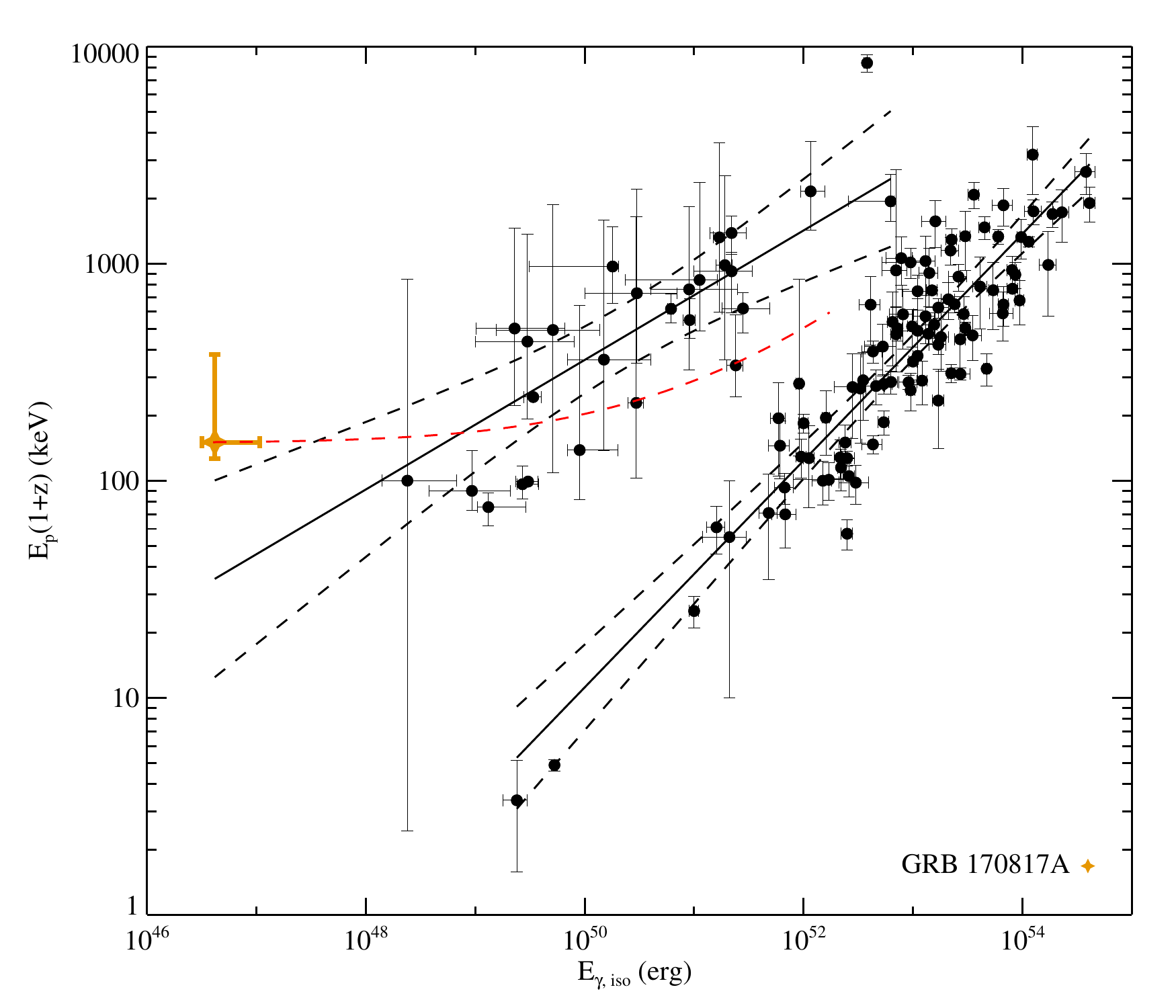} \\
\end{tabular}
\caption{GRB 170817A in the $E_p$-$E_{\rm iso}$ correlation diagram.The solid lines are the best fit correlations: log$E_{\mathrm{p}}/(1+z)=(3.24\pm0.07)+(0.54\pm0.04)$log$(E_{\mathrm{iso}}/10^{52})$ for short GRBs, log$E_{\mathrm{p}}/(1+z)=(2.22\pm0.03)+(0.47\pm0.03)$log$(E_{\mathrm{iso}}/10^{52})$ for long GRBs. Red dashed line represents GRB 170817A position if it were in different redshifts ranging from 0.009 to 3. All error bars represent 1-$\sigma$ uncertainties.}
\label{fig:epeiso}
\end{figure}

We next investigate the intrinsic property of the burst. Taking into the very small distance $D_{\rm L} \sim 40$ Mpc
of the host galaxy NGC 4993 \cite{2008ApJ...675.1459K}, this burst is abnormally low in terms of luminosity and energy (throughout the paper, luminosity and energy are the isotropic-equivalent ones). {The peak isotropic luminosity {with 50 ms bin size} is $1.6_{-0.4}^{+2.5} \times 10^{47} \ {\rm erg \ s^{-1}}$, and the isotropic energy is $E_{\rm iso} = 4.17_{-0.99}^{+6.54} \times 10^{46} $ erg}. Such low-luminosity sGRBs have never been observed before. Plotting it onto the intrinsic peak energy $E_{p,z} = E_p (1+z)$ vs. isotropic energy $E_{\rm iso}$ plane \cite{amati02,zhangb2009}, {we find that it is within the 2$\sigma$ of the track of the sGRB population, but slightly deviates from the 1$\sigma$ region of the track into the hard regime even if its $E_p$ error in included} (Figure \ref{fig:epeiso}). The burst would be more normal if its isotropic luminosity were somewhat higher.

\begin{figure}

\begin{tabular}{cc}
\includegraphics[keepaspectratio, clip, width=0.60\textwidth]{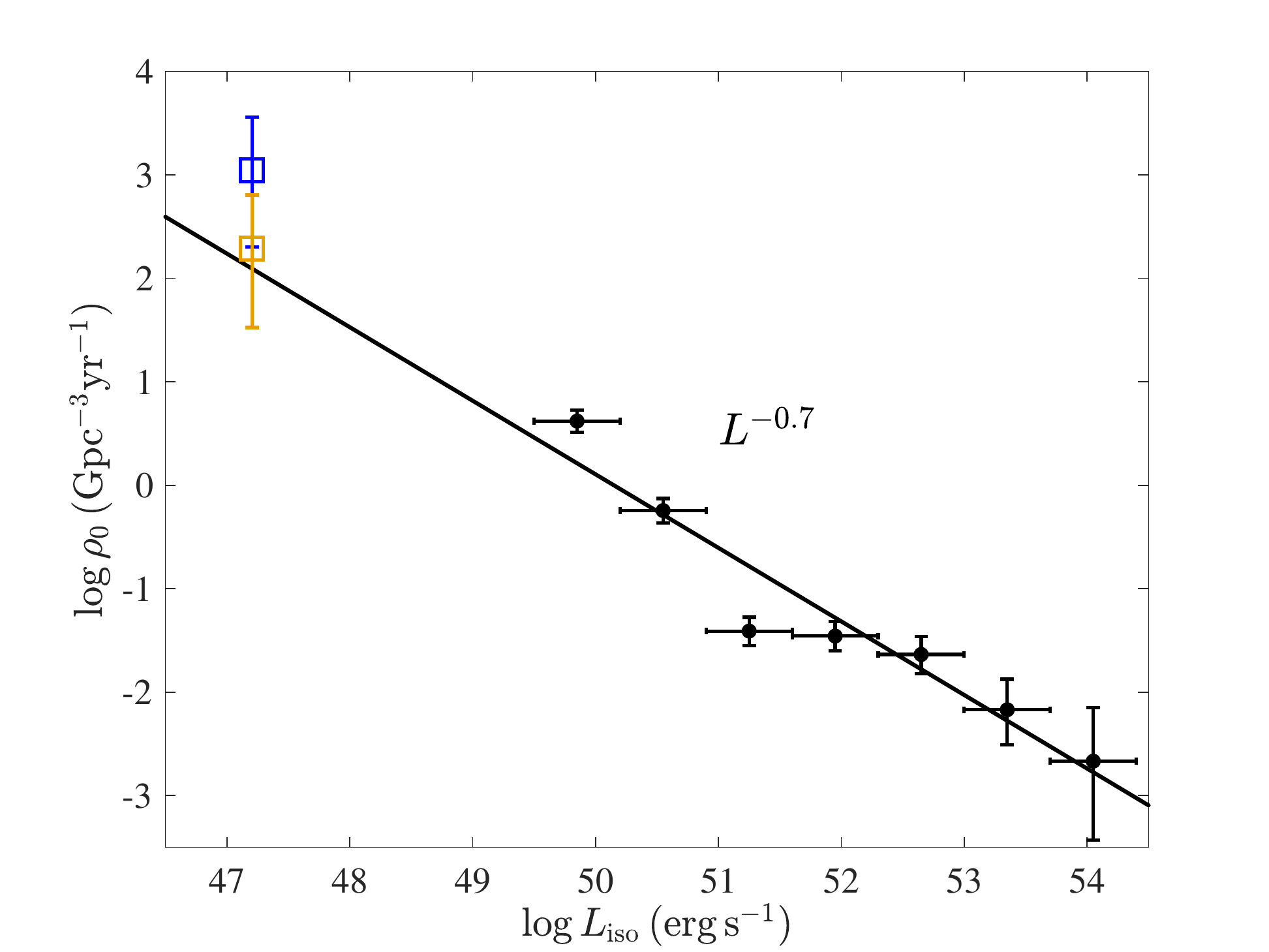} \\
\includegraphics[keepaspectratio, clip, width=0.60\textwidth]{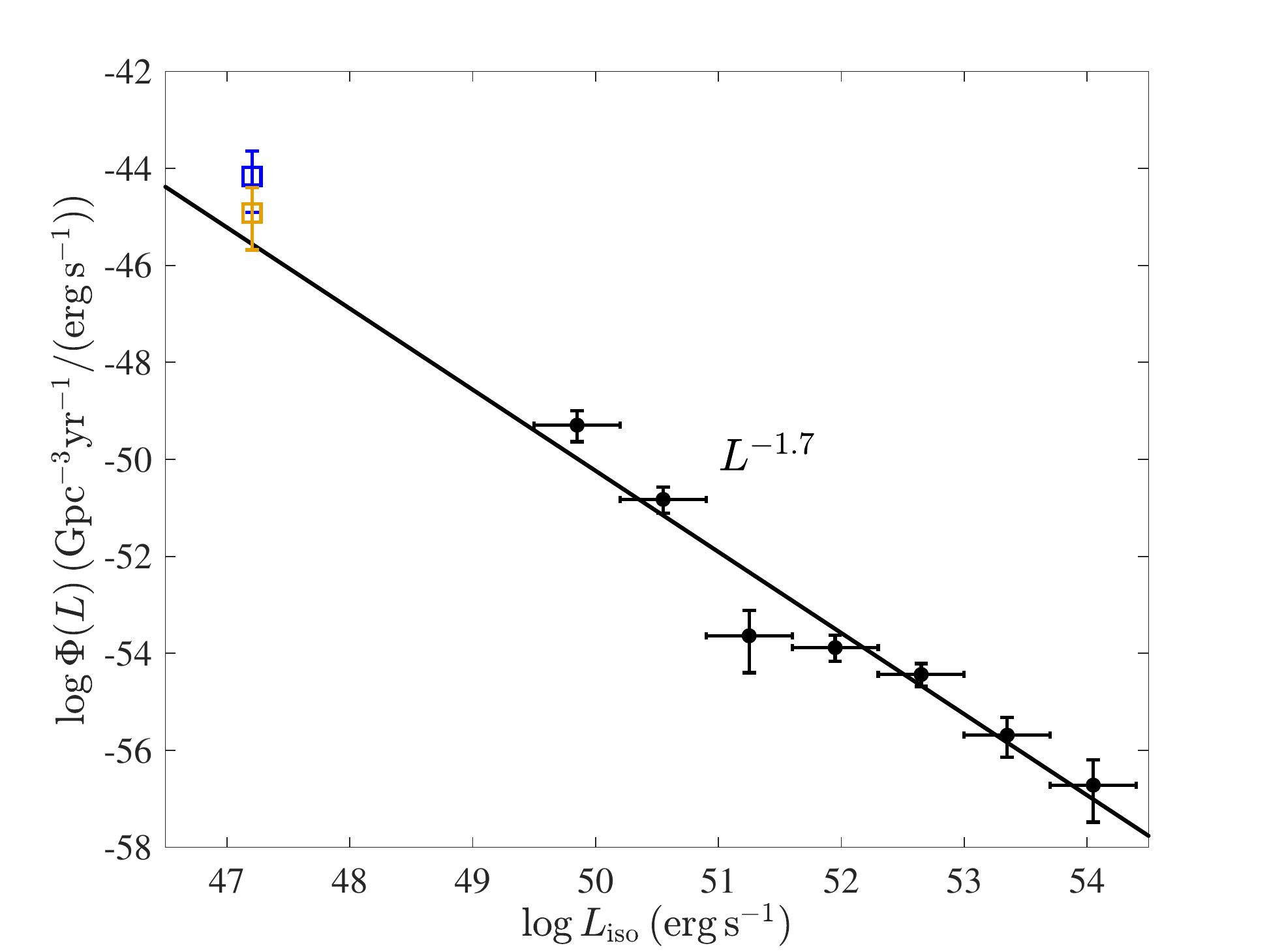} \\
\end{tabular}
\caption{Distributions of local event rate density and luminosity function. {\it a:} The local event rate density distribution of sGRBs including GRB 170817A. {The orange symbol with error denotes the event rate density derived from GRB 170817A} and the blue symbol with error denotes the DNS merger event rate density derived from GW170817. The black power-law line and other data points were derived from the {\em Swift} sGRB sample \cite{sun15}. The vertical error bar represents the $1\sigma$ Gaussian errors {derived from \cite{gehrels86}}.
{\it b:} Luminosity function distribution of sGRBs including GRB 170817A, with labels same as the upper panel. All error bars represent 1-$\sigma$ uncertainties.}
\label{fig:event rate}
\end{figure}

\noindent\textbf{Event rate density of the 170817A-like GRBs.}
Based on previously known sGRBs, the event rate density (also called volumetric event rate) of sGRBs above then-minimum luminosity ($\sim 10^{50}$ $\rm erg\,s^{-1}$) is a few $\rm Gpc^{-3}yr^{-1}$ \cite{wanderman15,sun15}. For example, for a Gaussian distribution of the merger delay time \cite{virgili11}, the event rate density of sGRBs is $4.2^{+1.3}_{-1.0}$ $\rm Gpc^{-3}yr^{-1}$ above $7\times 10^{49}$ $\rm erg\,s^{-1}$ \cite{sun15}. This was significantly lower than the estimated DNS merger event rate density (\cite{prl}, see below).
The discrepancy may be removed if one considers the beaming correction of sGRBs {within the top-hat uniform jet model}. Using the beaming factor $f_b \sim 0.04$ inferred from the sparse sGRB jet break data collected in the past \cite{fong2015}, the beaming-corrected event rate density (counting for sGRBs not beaming towards us) was $\sim 100$ $\rm Gpc^{-3}yr^{-1}$. With the detection of GRB 170817A, the distribution of the sGRB isotropic peak luminosity extended down by $\sim$ three orders of magnitude. The revised event rate density of sGRB above $1.6 \times 10^{47}$ $\rm erg\,s^{-1}$ becomes (Methods)
\begin{equation}
\rho_{\rm 0,sGRB} (L_{\rm iso}>1.6 \times 10^{47} { erg\,s^{-1}}) = 190^{+440}_{-160} \ {Gpc^{-3} \, yr^{-1}}
\end{equation}
{if one assumes only one such event within the GBM archives.}
This is comparable to (or somewhat higher than) the previously-derived beaming-corrected sGRB event rate density, {but could be still up to a factor of a few} smaller than the DNS merger event rate density derived based on the detection of GW 170817A \cite{prl}, which is (Methods)
\begin{equation}
\rho_{\rm 0,DNS} = 1100^{+2500}_{-910} \ { Gpc^{-3} \, yr^{-1}}.
\end{equation}
Figure \ref{fig:event rate} upper panel shows the sGRB event rate density as a function of luminosity threshold. The black power-law (PL) line with an index $-0.7$ was derived from the {\em Swift} sGRBs (black crosses with error included, {peak luminosity derived with 64 ms time bin}) with redshift measurements \citep{sun15}. 
GRB 170817A (orange) extends the sGRB luminosity by three orders of magnitude in the low-$L_{\rm iso}$ regime. Interestingly, the revised event rate density above $1.6\times 10^{47} \ {\rm erg \ s^{-1}}$ follows the extension of the PL distribution derived by \cite{sun15}. {If one considers that there might be some sGRBs similar to GRB 170817A hidden in the GBM archives, the true event rate density could be higher, but has to be limited by the DNS merger event rate density (blue symbol)}. 
In Figure \ref{fig:event rate} lower panel, we derive a new sGRB luminosity function across a wide range of luminosity, which is consistent with the extrapolation of the previous results that show a power law with $L_{\rm iso}^{-1.7}$ \cite{sun15}.

\section{Discussion}

There are in principle two possibilities to produce a low-luminosity sGRB from an DNS merger. The first possibility is a bright sGRB jet viewed off-axis. Within this picture, the main jet (similar to the one observed from a more distant sGRB) beams towards a different direction. However, within such a scenario, one cannot have a sharp-edge conical jet viewed outside the jet cone. This is because the observed duration would be longer than the central engine activity time scale, inconsistent with its typical sGRB duration (Supplementary Note 6). Rather, one requires a structured jet viewed from a large wing \cite{zhangmeszaros02,rossi02} with emission powered by the low-luminosity wind along the line of sight. Within the sGRB context, such a jet configuration has been discussed in terms of a jet-cocoon geometry \cite{lamb17,lazzati17}. {A viewing angle $\theta_v \leq 28^{\rm o}$ (or $\leq 36^{\rm o}$ depending on the assumed value of the Hubble constant \cite{ligoapjl}) has been inferred from the gravitational wave data. This is consistent with such a scenario. 
{The second possibility is that the outflow of GRB 170817A may have an intrinsically low luminosity. However, the late rise of X-ray and radio flux \cite{troja17,hallinan17} from the source suggests that the total energy budget of the source is higher. It disfavors this second possibility but favors the off-axis structured jet scenario (Supplementary Note 9).}

{The short duration of the burst is consistent with a prompt black hole or a hyper-massive neutron star that survived for a short (e.g. $\sim$ 100 ms) period of time before collapsing to a black hole. 
We conduct a search of precursor or extended emission before and after the GRB trigger time and give a negative result (Supplementary Note 5). Even though the possibility of a long-live post-merger neutron star product cannot be ruled out from the GW \cite{LIGOproduct} and EM data, our non-detection of extended $\gamma$-ray emission is consistent with a BH post-merger product (Supplementary Note 7).}

{The merger time of the gravitational wave signal is at T$_{GW}$=12:41:04.430$\pm$0.002 UTC on 17 August 2017 (GPS time T$_{GW}$= 1187008882.430$\pm$0.002 s)\cite{ligoapjl}. The beginning time of GRB 170817A ($\sim -0.3$ s with respect to the {\it Fermi}/GBM trigger time $T_0$=12:41:06.474598 UT on 17 August 2017\cite{gbm_apjl}) has a $\sim 1.7$ s delay with respect to the merger time.} It is intriguing that this delay time scale has the same order as the burst duration itself. Such a delay offers a diagnostic of the emission site and energy dissipation process of GRBs. {In particular, a scenario that invokes a magnetized jet dissipating in an optically thin region can interpret both time scales simultaneously without introducing an ad hoc jet-launching delay time as most other models do (Supplementary Note 8). }

Assuming a standard radiative efficiency and standard shock microphysics parameters, the low isotropic energy of GRB 170817A suggests that the multi-wavelength afterglows of the burst should be very faint (Supplementary Note 6). 
We used the Javier Gorosabel 0.6m robotic telescope at the BOOTES-5
station at Observatorio Nacional de San Pedro Martir (Mexico) to image
the 15 galaxies in the GLADE Catalogue starting on Aug 18.21 UT. 
The optical counterpart (SSS17a) of GW 170817 was detected in the outskirts of the NGC 4993
galaxy, with a magnitude
R = 18.20 $\pm$ 0.45, in agreement with other contemporaneous
measurements. This is much brighter than the predicted flux of optical afterglow. As a result, this optical transient originates from a quasi-thermal kilonova \cite{li98,metzger10}, {as suggested by independent modeling of many authors (e.g. \cite{kasen17}). }

Within the Fermi GBM soft faint sGRB sample, there might be at most GRB 170817A-like events limited by the DNS merger rate. Some short, faint sGRB events are presented in Supplementary Figure 6. However, identifying them turns out difficult without gravitational wave detections (Supplementary Note 10).

\vspace{\baselineskip}

\section{Methods}

\noindent\textbf{Determine GRB Duration Using 2-D Energy vs. Time Count Map.} {The GRB duration is usually defined by T$_{90}$, which is the time span over which 5\% to 95\% of its total measured counts are measured \cite{GBM}. The calculation of T$_{90}$ is subject to the selection of energy band, the bin-size as well as the assumption of the model background (e.g. a 2nd or 3rd order polynomial function) of the GRB light curve. To minimize such an artificial effect for a faint GRB like GRB 170817A, we utilize the 2-D count map of the photon energy and photon arrival time directly from the {\it Fermi}/GBM Time-Tagged Event (TTE) data and calculate the GRB duration. Our procedure is the following: (1) select the source region between the time interval [t1, t2] that includes the GRB signal. For GRB 170817A, [t1,t2] = [-1,3]. (2) Select all the photon events between [t1,t2] in the {\it Fermi}/GBM Time-Tagged Event (TTE) data. Note the selected data are a list of [time, energy] pairs. (3) Convert time versus energy pairs to 2-D points in the time versus energy plot, then use Kernel Density Estimation (KDE) to plot a 2-D source count map in the time vs. energy plot. This is the top panel in Figure 1. (4) Select two background regions [t3, t4] and [t5,t6] that are before and after the burst region. For GRB 170817A, [t3, t4] = [-10,-2.] and [t5,t6]= [5,10]. Repeat steps (2) and (3) to get two 2-D count maps for the pre-burst and after-burst backgrounds. (5) Perform interpolation between those two background count maps to calculate the source-normalized background count map within [t1, t2] (middle panel of figure 2). Such a normalized and interpolated background within the source region can be used to calculate the standard derivation (STD) of the background. (6) Subtract the background count map from the source count map to get the net count map (bottom panel of Figure 1). For each bin in the count map, define its signal-to-noise ratio as S/N = ( net count ) / STD. Overplotting the S/N = 1, 3, 5 in the net count map, we then define the burst duration region as where S/N $\geq$ 5.0 is satisfied. }

\medskip
\noindent\textbf{sGRB event rate density.}
The abnormally low luminosity and extremely small distance of GRB 170817A suggest that the actual event rate density of short GRBs is large. With one detection, one can estimate the local event rate density $\rho_{\rm 0,sGRB}$ of short GRBs through
\begin{equation}
N_{\rm sGRB}=\frac{\Omega_{\rm GBM} T_{\rm GBM}}{4\pi}\rho_{\rm 0,sGRB}V_{\rm max} = 1.
\end{equation}
The field of view of GBM is approximatively taken as full sky with $\Omega_{\rm GBM}\simeq 4\pi$. The working time of GBM is taken since 2008 with a duty cycle of $\sim 50\%$, so that $T_{\rm GBM}\simeq 4.5 \rm yrs$. The maximum volume a telescope can detect for this low luminosity event is $V_{\rm max}=4\pi D_{\rm L,max}^3/3$. 
We simulate a set of pseudo-GRBs by placing GRB 170817A to progressively larger distances, and find that the signal would not be detectable at 65 Mpc (Supplementary Note 4; Supplementary Figure 8). Taking this distance as $D_{\rm L,max}$, we derive the event rate density of sGRBs \cite{sun15}
\begin{equation}
\rho_{\rm 0,sGRB} (L_{\rm iso}>1.6 \times 10^{47} {\rm erg\,s^{-1}}) = 190^{+440}_{-160} \ {\rm Gpc^{-3} \, yr^{-1}}
\label{eq:sgrb}
\end{equation}
{assuming only one such sGRB exists in the GBM archives. This number may be regarded as a lower limit if in reality there are other hidden ones.}

The event rate density of DNS mergers may be also estimated based on one detection by aLIGO during O1 and O2. 
Since only one DNS merger event was detected \cite{prl}, one may write
\begin{equation}
N_{\rm DNS}=\frac{\Omega_{\rm LVC}}{4\pi}\rho_{\rm 0,DNS} (V_{\rm max,O1} T_{\rm O1} + V_{\rm max,O2} T_{\rm O2}) = 1.
\end{equation}
Noticing $\Omega = 4\pi$ for GW detectors, taking DNS merger horizon $\sim 60$ Mpc and $\sim 80$ Mpc for O1 and O2, respectively, and adopting a duty cycle of $\sim 40\%$ for both O1 and O2, we estimate
\begin{equation}
\rho_{\rm 0,DNS} = 1100^{+2500}_{-910} \ {\rm Gpc^{-3} \, yr^{-1}}.
\label{eq:nsns}
\end{equation}
This is consistent with the DNS merger event rate density derived by the LIGO-VIRGO team using more sophisticated simulations \cite{prl}.

{The error bars in both Equations \ref{eq:sgrb} and \ref{eq:nsns} show the $1\sigma$ Gaussian errors derived from \cite{gehrels86} by taking only one observational event into account. Comparing the two equations, one can see that even though the sGRB rate density may be consistent with the DNS merger rate density, it can be smaller than the latter by up to a factor of a few. 
This either suggests that there might be even less luminous sGRBs than GRB 170817A, or there might be GRB 170817A-like sGRBs hidden in the GBM archives. The number of these events is at most a few. }

\vspace{\baselineskip}
\textbf{References}

\makeatletter
\def\@biblabel#1{\@ifnotempty{#1}{#1.}}

\vspace{\baselineskip}

\noindent\textbf{Acknowledgments}

We acknowledge support from National Basic Research Program of China (973 Program, Grant No. 2014CB845800). BBZ and AJCT acknowledge support from the Spanish Ministry Projects AYA 2012-39727-C03-01 and AYA2015-71718-R.  BBZ acknowledge the support from the National Thousand Young Talents program of China. This work is also supported by the Strategic Pioneer Program on Space Science, Chinese Academy of Sciences (Grant No.XDA15052100).  BZ, HS, HG and XFW are supported by the Strategic Priority Research Program of the Chinese Academy of Sciences ``Multi-waveband Gravitational Wave Universe"(Grant No. XDB23040000). HG acknowledges the National Natural Science Foundation of China under grants No. 11722324, 11603003, 11633001 and 11690024. HJL acknowledges the National Natural Science Foundation of China under grants 11603006. YDH acknowledges the support by China Scholarships Council (CSC) under the Grant No.201406660015. XYW acknowledges the National Natural Science Foundation of China under grant 11625312. LS acknowledges
the supported by the Joint NSFC-ISF Research Program (No.
11361140349), jointly funded by the National Natural Science
Foundation of China and the Israel Science Foundation. XF was supported by Natural Science Foundation of China under Grants No. 11673008. The BOOTES-5/JGT observations were carried out at observatories
Astronomico Nacional in San Pedro Martir (OAN-SPM, Mexico), operated by
Instituto de Astronomia, UNAM and with support from Consejo Nacional de
Ciencia y Tecnologia (Mexico) through the Laboratorios Nacionales
Program (Mexico), Instituto de Astrofisica de Andalucia (IAA-CSIC,
Spain) and Sungkyunkwan University (SKKU, South Korea). We thank the
staff of OAN-SPM for their support in carrying out the observations. We acknowledge the use of the public data from the {\it Fermi} and {\it INTEGRAL} data archives.

\vspace{\baselineskip}

\vspace{\baselineskip}
\cleardoublepage


\makeatletter
\def\@biblabel#1{\@ifnotempty{#1}{#1.}}
\apptocmd{\thebibliography}{\global\c@NAT@ctr 35\relax}{}{}

\newpage
\clearpage
\beginsupplement

\vspace{\baselineskip}

\vspace{\baselineskip}
\textbf{Supplementary Information}

\leftline{\textbf{Supplementary Note 1. Detailed light curves and spectral fit to GRB 170817A}}

The multi-channel light curves observed by Fermi/GBM and INTEGRAL/SPI-ACS is presented in Supplementary Figure \ref{fig:multilc}.

As discussed in the main text, a cutoff power law presents an adequate fit to the spectral data. The best fitting result for the time interval (T$_0$-0.26,T$_0$+0.57) is presented in Supplementary Figure \ref{fig:fitall}. We noticed that a simple power law model can also fit the data with a power law index -1.61$_{-0.13}^{+0.09}$ and PGSTAT/dof 261.9/352. To check whether the cutoff power law fit is overfitting (since it has one extra parameter), we employ a Bayesian information criterion (BIC) \cite{Kass95} to check its statistical confidence. As shown in Supplementary Table \ref{tab:specfitting}, the comparison of the two models leads to $\Delta$BIC =3.38 (cutoff power law model has the lower BIC). As suggested by \cite{Kass95}, such a $\Delta$BIC value indicates positive evidence against the model with a higher BIC (Power Law model in this case). So we favorably choose the cutoff power law model throughout our analysis.

We also compare the fits between the blackbody (BB) model and the CPL model. For the time-integrated spectral fitting, according to BIC, the BB model is less preferred to fit the observed spectra. We also notice the best-fit low energy photon index $-0.66$ in the CPL model is too soft to match the blackbody value (+1). 

{According to BIC, the weaker emission between 0.95 s and 1.79 s is favorably fitted by the BB model with $kT = 11.3_{-2.36}^{+3.85}$ keV. The best fitting result of this time interval is presented in Supplementary Figure \ref{fig:fit_taill}. }

\begin{figure}
\begin{tabular}{c}
\includegraphics[keepaspectratio, clip, width=0.85\textwidth]{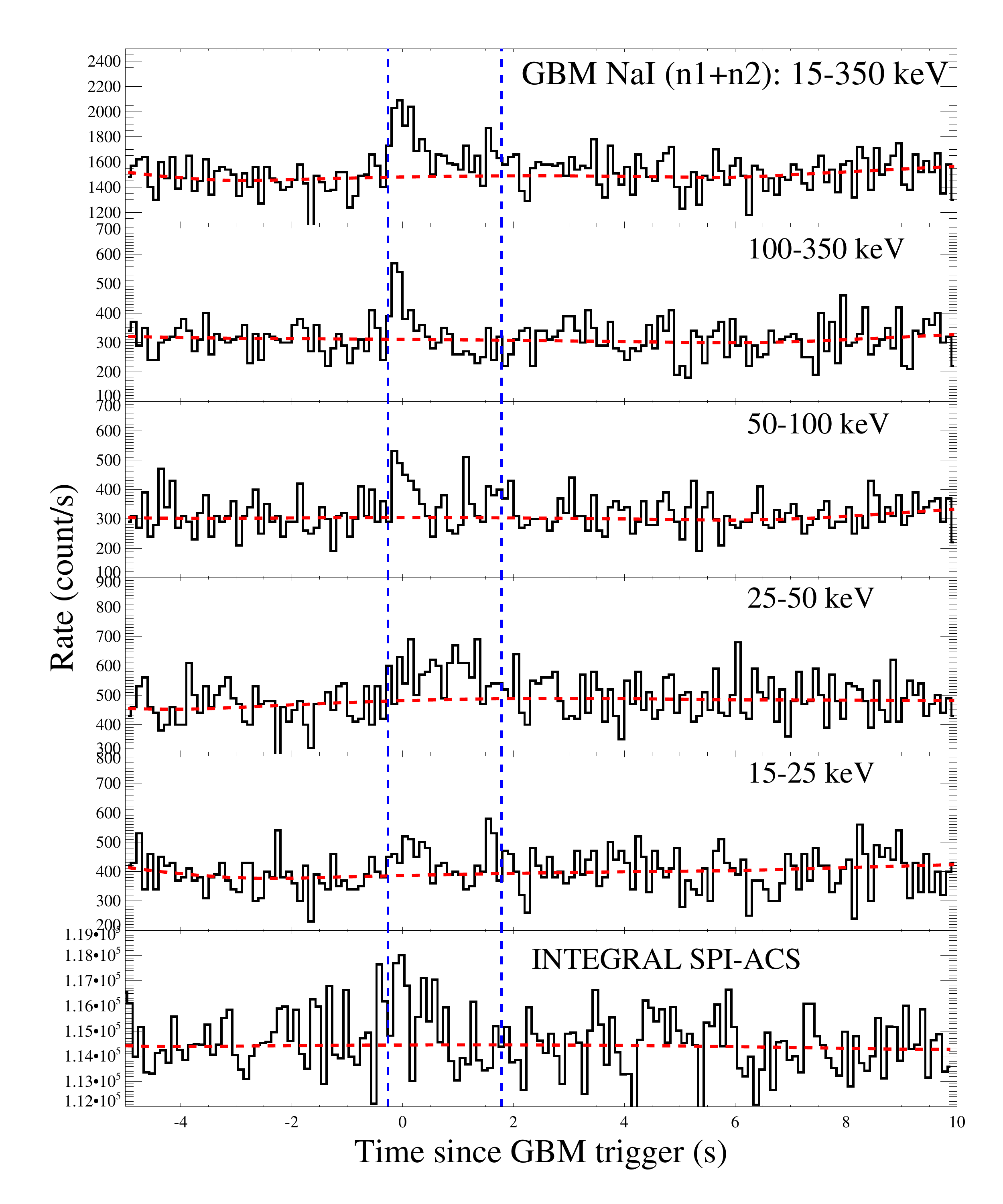} \\
\end{tabular}
\caption{Multi-Channel Light Curves observed by {\it Fermi}/GBM and {\it INTEGRAL}/SPI-ACS. The two vertical dashed lines indicate the S/N>5 region as shown in Figure 1. }
\label{fig:multilc}
\end{figure}

\begin{figure}

\begin{tabular}{c}
\includegraphics[keepaspectratio, clip, width=0.35\textwidth]{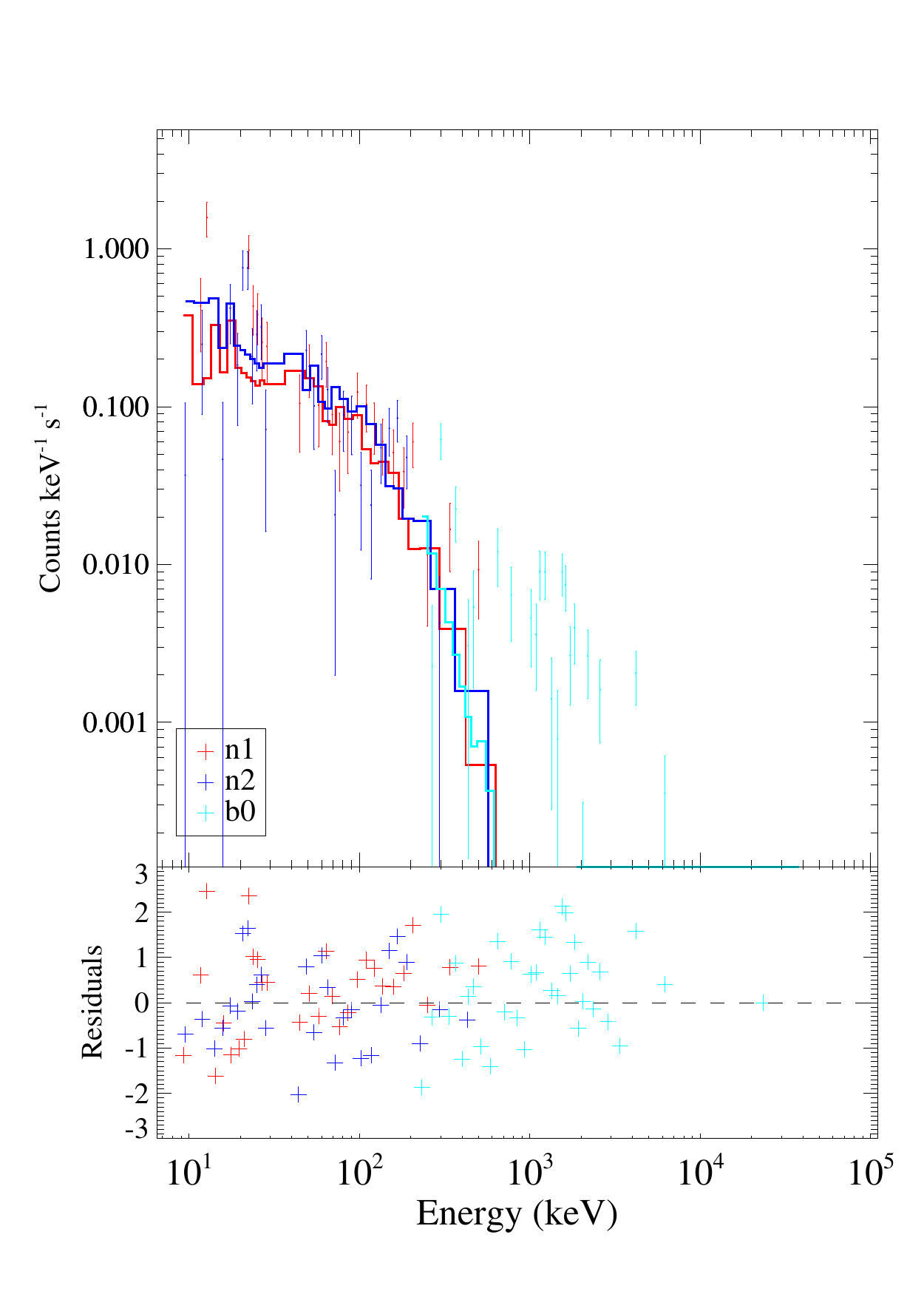} \\
\includegraphics[keepaspectratio, clip, width=0.35\textwidth]{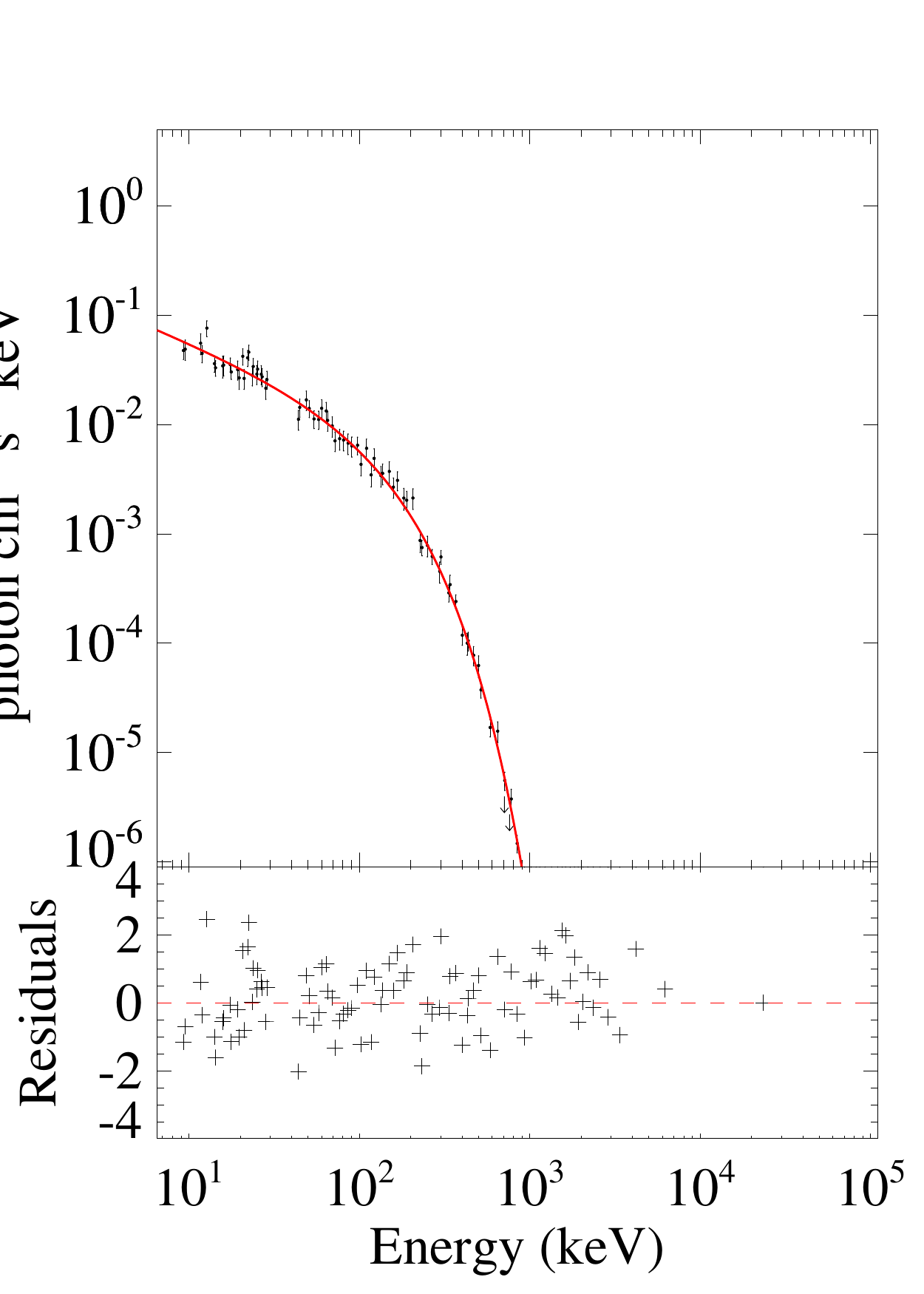} \\
\includegraphics[keepaspectratio, clip, width=0.35\textwidth]{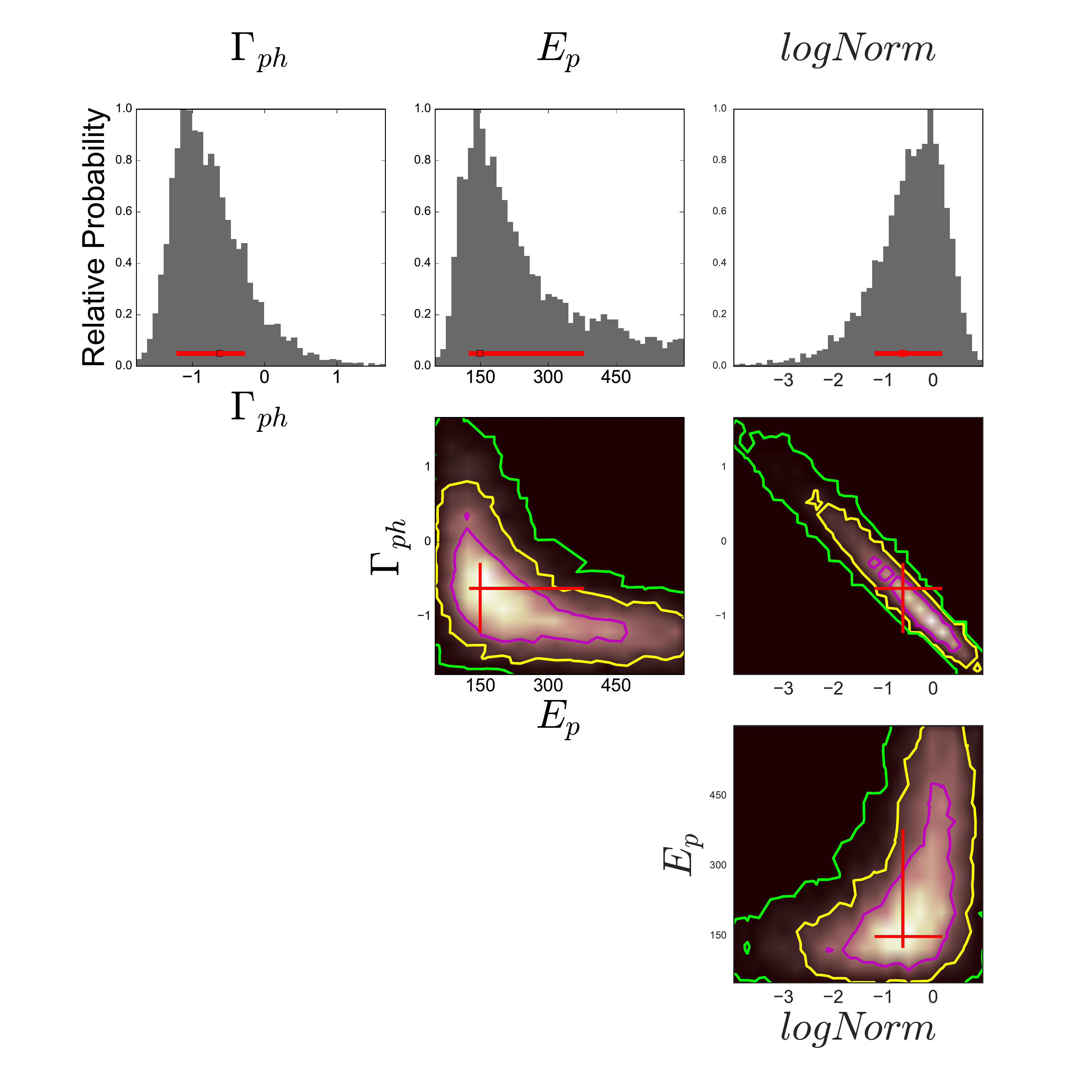} \\
\end{tabular}
\caption{Spectral fitting result for the interval ($T_0-0.26$, $T_0+0.57$). {\it a}: count spectrum. {\it b:} de-convolved photon spectrum. {\it c}: parameter likelihood map. All error bars represent 1-$\sigma$ uncertainties. }
\label{fig:fitall}
\end{figure}

\begin{table}

\caption{\label{tab:specfitting}Spectral Fitting Results of GRB 170817A}
\setlength{\tabcolsep}{2pt}

\begin{tiny}
\begin{center}
\begin{tabular}{c|cccc|ccc|ccc|cc}
\hline

Time & \multicolumn{4}{|c}{Cutoff Power-Law Fitting} & \multicolumn{3}{|c|}{Power-Law Fitting} & \multicolumn{3}{|c|}{Blackbody Fitting} & \multicolumn{2}{c}{Model Comparison} \\
\hline
 t1 $\sim$ t2 (s) & $\alpha$ & E$_p$ (keV) & $\frac{\textrm PGSTAT}{\textrm dof}$ & BIC & $\alpha$ & $\frac{\textrm PGSTAT}{\textrm dof}$& BIC & kT (keV) & $\frac{\textrm PGSTAT}{\textrm dof}$ &BIC &
 BIC$_{pl}$ - BIC$_{cpl}$ & BIC$_{bb}$ - BIC$_{cpl}$ \\
\hline

-0.26 $\sim$ 0.57 & -0.61$_{-0.60}^{+0.34}$ & 149.1$_{-24.2}^{+229.4}$ & 252.7/351 & 270.29 & -1.61$_{-0.13}^{+0.09}$ & 261.9/352 & 273.68& 29.0 $_{-8.5}^{+10.5}$ & 261.1/352 & 272.83& 
	3.38 &2.53 \\
	 		


-0.3 $\sim$ 0.05 & 0.07$_{-0.92}^{+0.72}$ & 147.9$_{-28.4}^{+160.7}$& 237.1/352 & 254.70& -1.65$_{-0.44}^{+0.12}$ & 248.0/352 & 259.71& 34.6 $_{-7.3}^{+13.6}$ & 238.2/352 & 249.89 &
	 5.01 & -4.8 \\



0.05 $\sim$ 0.4 & -0.78$_{-0.84}^{+1.01}$ & 62.4$_{-21.8}^{+77.5}$ &209.1/351 &226.77 
& -2.12$_{-0.52}^{+0.22}$ & 211.9/352 &223.66 
& 10.5 $_{-2.1}^{+12.2}$ & 211.2/352 & 222.97&
 -3.11 & -3.79\\


0.95 $\sim$ 1.79 &2.65$_{-3.66}^{+0.23}$ & 42.5$_{-11.1}^{+28.2}$ & 236.3/351 &253.90 
 & -2.27$_{-2.86}^{+0.63}$ & 241.4 /352 &253.19 
& 11.3 $_{-2.36}^{+3.85}$ & 236.35/352 & 248.09&
	-0.72	& -5.81 \\ 
		

\hline
\end{tabular}
 \end{center}
\end{tiny}
\end{table}


%
%

\begin{figure}

\begin{tabular}{c}
\includegraphics[keepaspectratio, clip, width=0.35\textwidth]{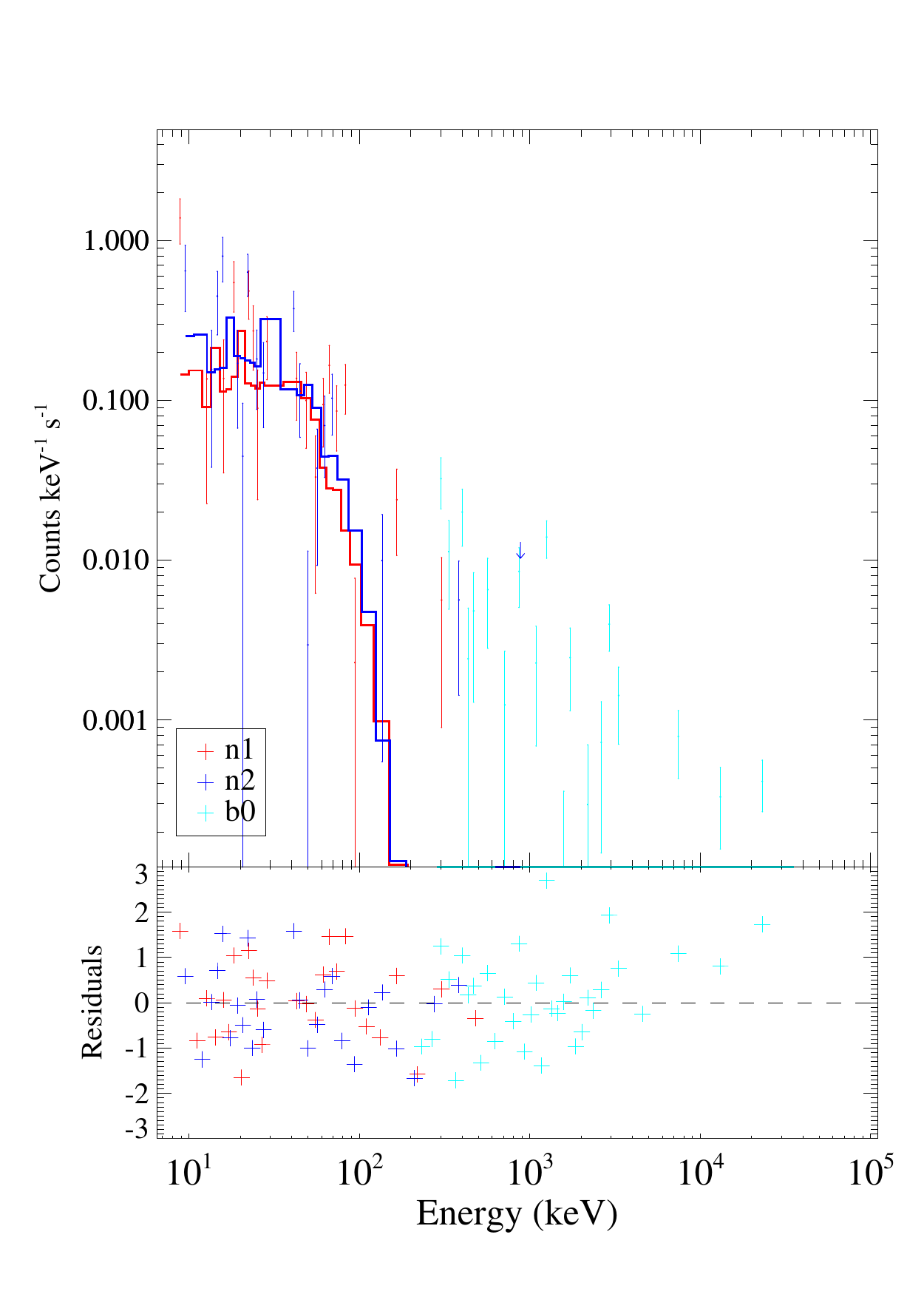} \\
\includegraphics[keepaspectratio, clip, width=0.35\textwidth]{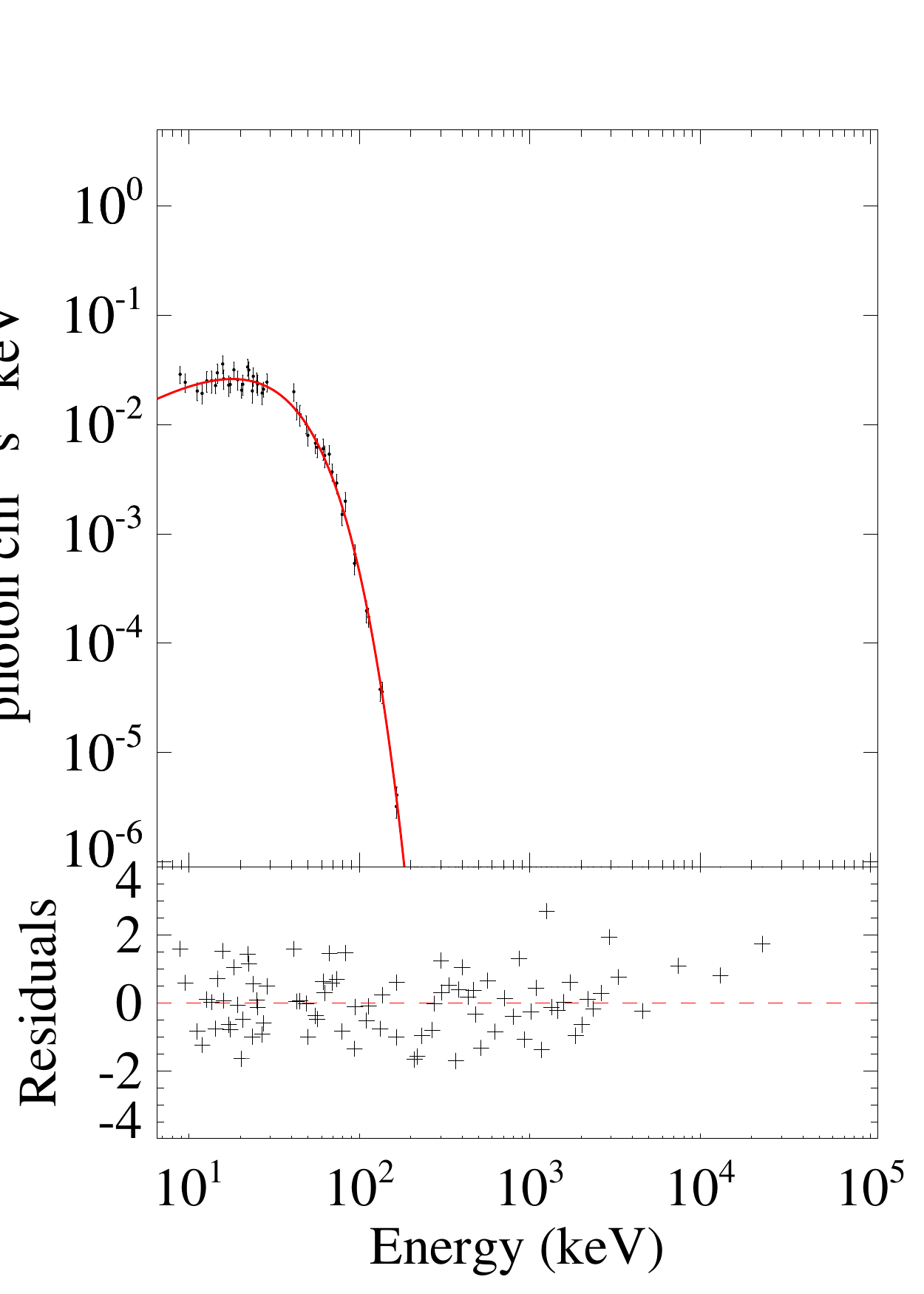} \\
\includegraphics[keepaspectratio, clip, width=0.35\textwidth]{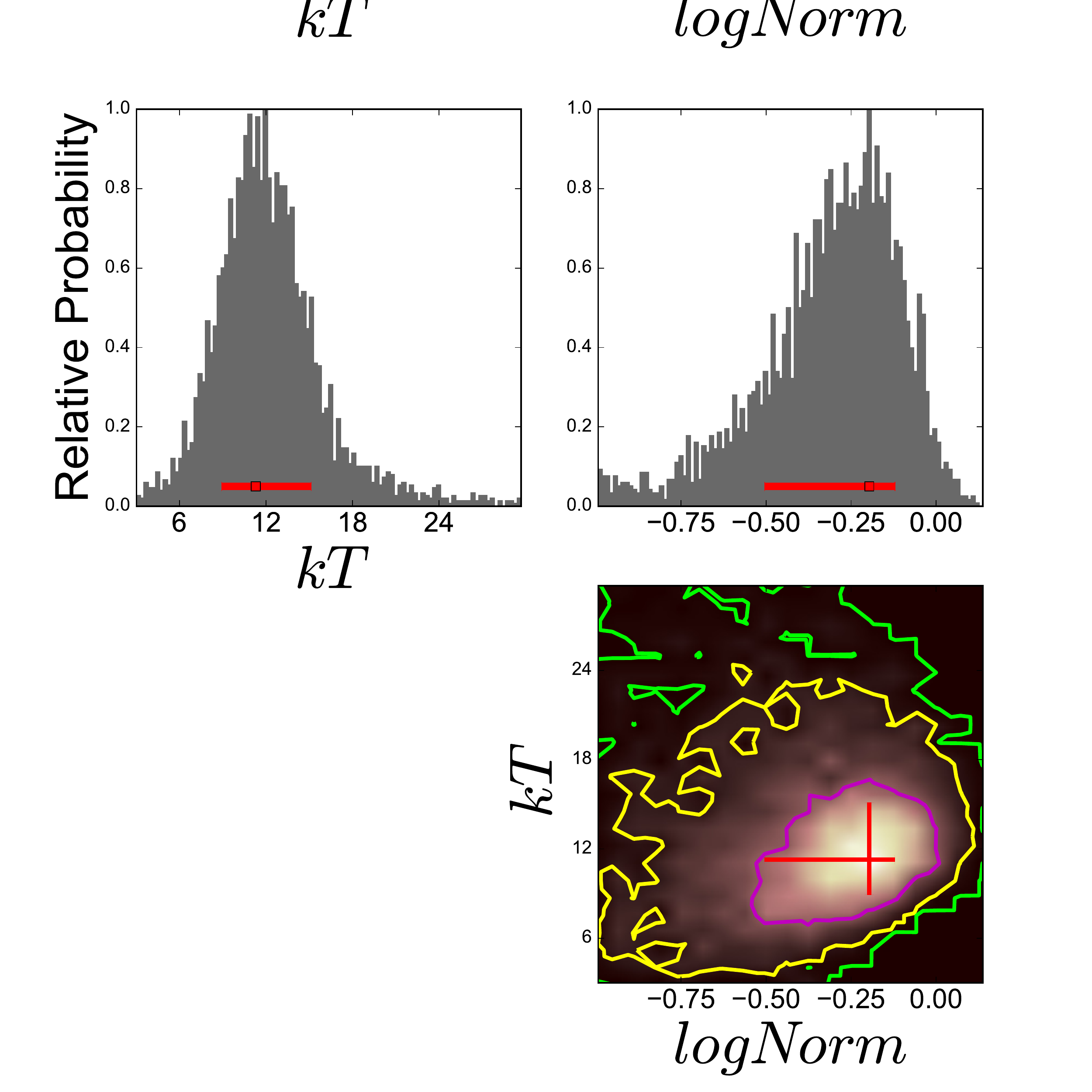} \\
\end{tabular}
\caption{Spectral fitting result for the interval ($T_0+0.95$, $T_0+1.79$) s. {\it a}: count spectrum. {\it b}: de-convolved photon spectrum. {\it c}: parameter likelihood map. All error bars represent 1-$\sigma$ uncertainties.}
\label{fig:fit_taill}
\end{figure}

Due to the low number of photon counts and short duration, the finest bin size we are able to perform a time-resolved spectral analysis is around 0.2-0.3 s, below which the spectral parameters become unconstrained. We select the brightest region between $T_0-0.3$ s and $T_0+0.4$ s in the first peak, divide it into two equal slices, and perform the time-dependent spectral analysis on them. We present the spectral evolution properties in Supplementary Table \ref{tab:specfitting} and Supplementary Figure \ref{fig:spec_evo}. Our analysis suggests that there are indeed some spectral differences between the two slices. 
 %

\begin{figure}

\begin{tabular}{c}
\includegraphics[keepaspectratio, clip, width=0.85\textwidth]{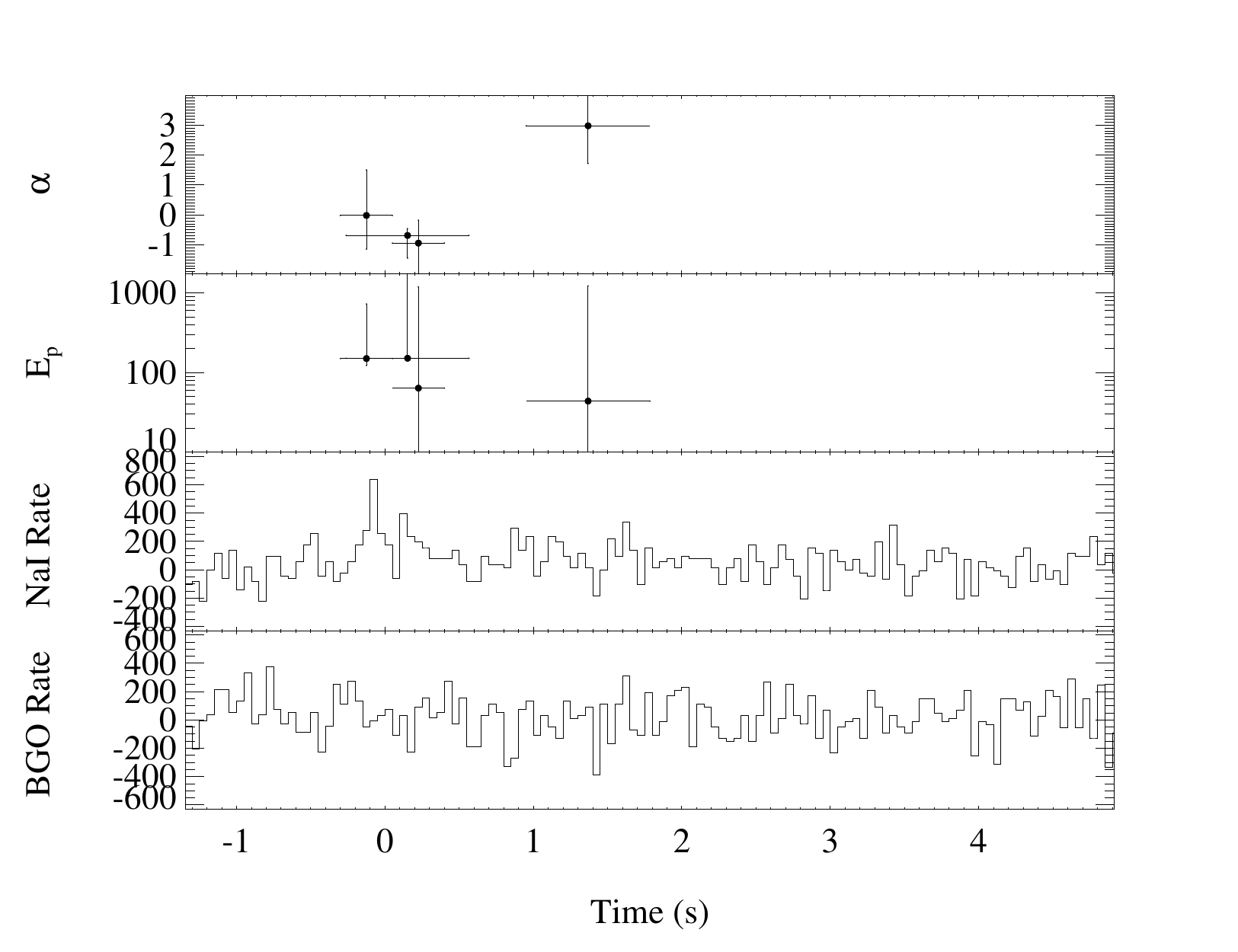} \\
\end{tabular}
\caption{Spectral evolution of the burst. Top two panels show evolution of the photon index ($\alpha$ ) and peak energy (E$_p$). The time intervals are listed in Supplementary Table 1. Bottom two panels show the NaI (n1) light curve in 15-350 keV and BGO light curve in 250- 20000 keV. All error bars represent 1-$\sigma$ uncertainties.}
\label{fig:spec_evo}
\end{figure}

\leftline{\bf Supplementary Note 2. Definition of The Faint Short GRB Sample }

In Supplementary Figure 6, we plot all the Fermi GBM short GRBs in terms of their signal-to-noise ratio (S/N), fluence, and $E_p$. The bin where GRB 170817A is located in is marked as the red dashed vertical line. The sample to the left of the line in the S/N plot is defined as the faint sGRB sample. We plot some examples of the light curves of this sample in Supplementary Figure \ref{fig:weak-bursts}.

\begin{figure}

\begin{tabular}{c}
\includegraphics[keepaspectratio, clip, width=0.55\textwidth]{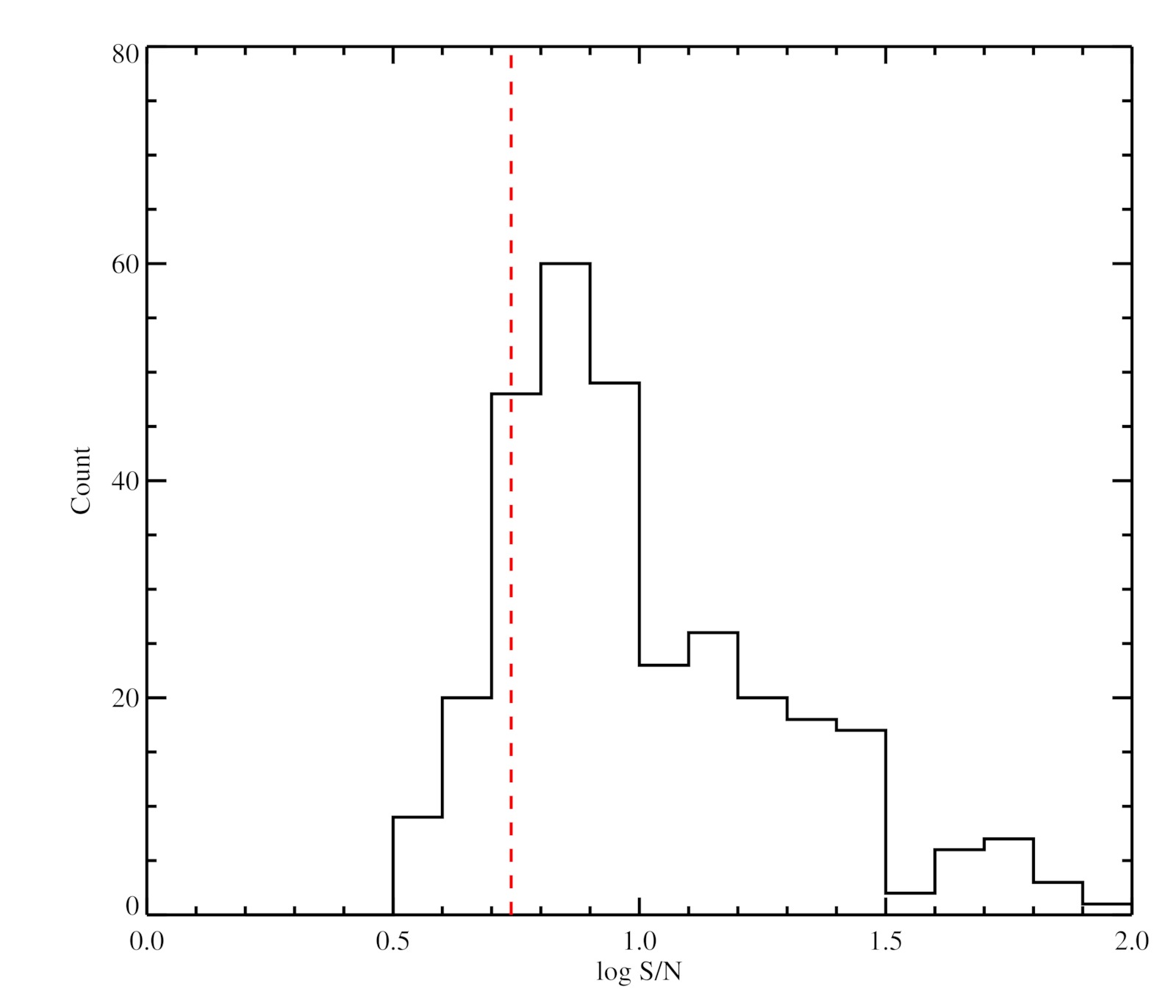} \\
\includegraphics[keepaspectratio, clip, width=0.55\textwidth]{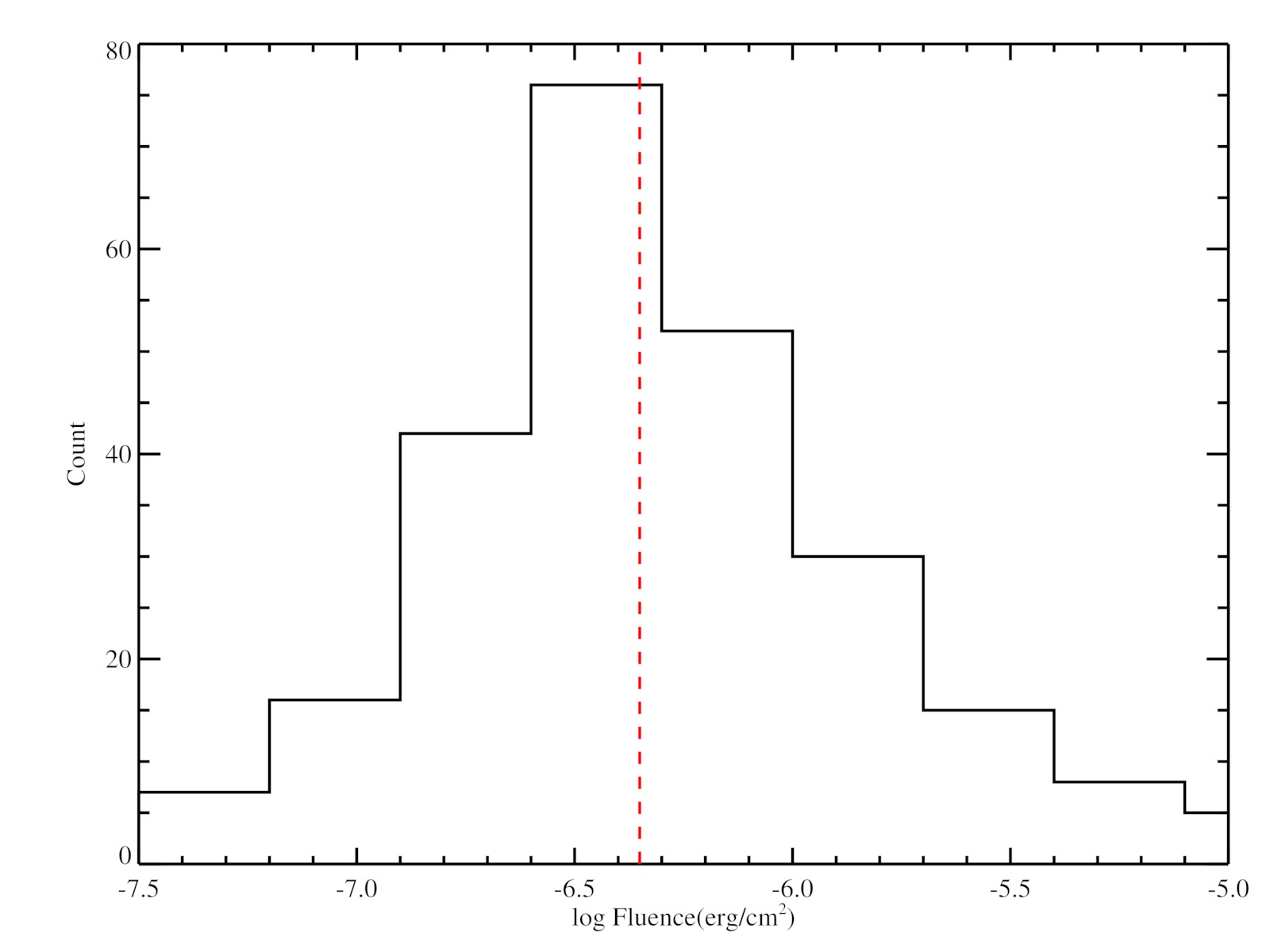} \\
\includegraphics[keepaspectratio, clip, width=0.55\textwidth]{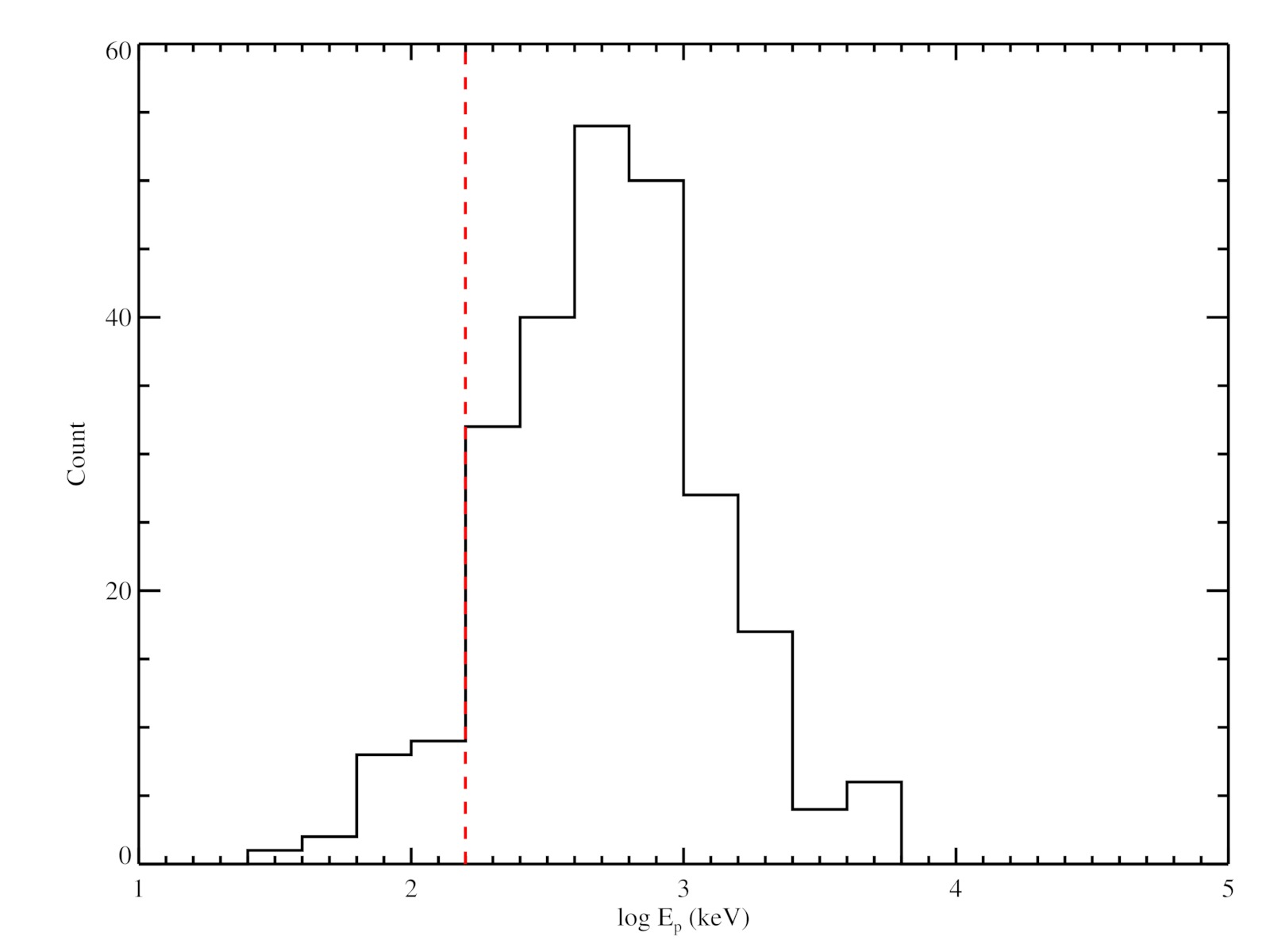} \\

\end{tabular}
\caption{The Fermi GBM short GRB histograms in terms of S/N ratio distribution, fluence distribution, and $E_p$ distribution. The vertical red lines indicate the values for GRB 170817A.}
\label{fig:fluence_ep}
\end{figure}

\begin{figure}

\begin{tabular}{c}
\includegraphics[keepaspectratio, clip, width=0.95\textwidth]{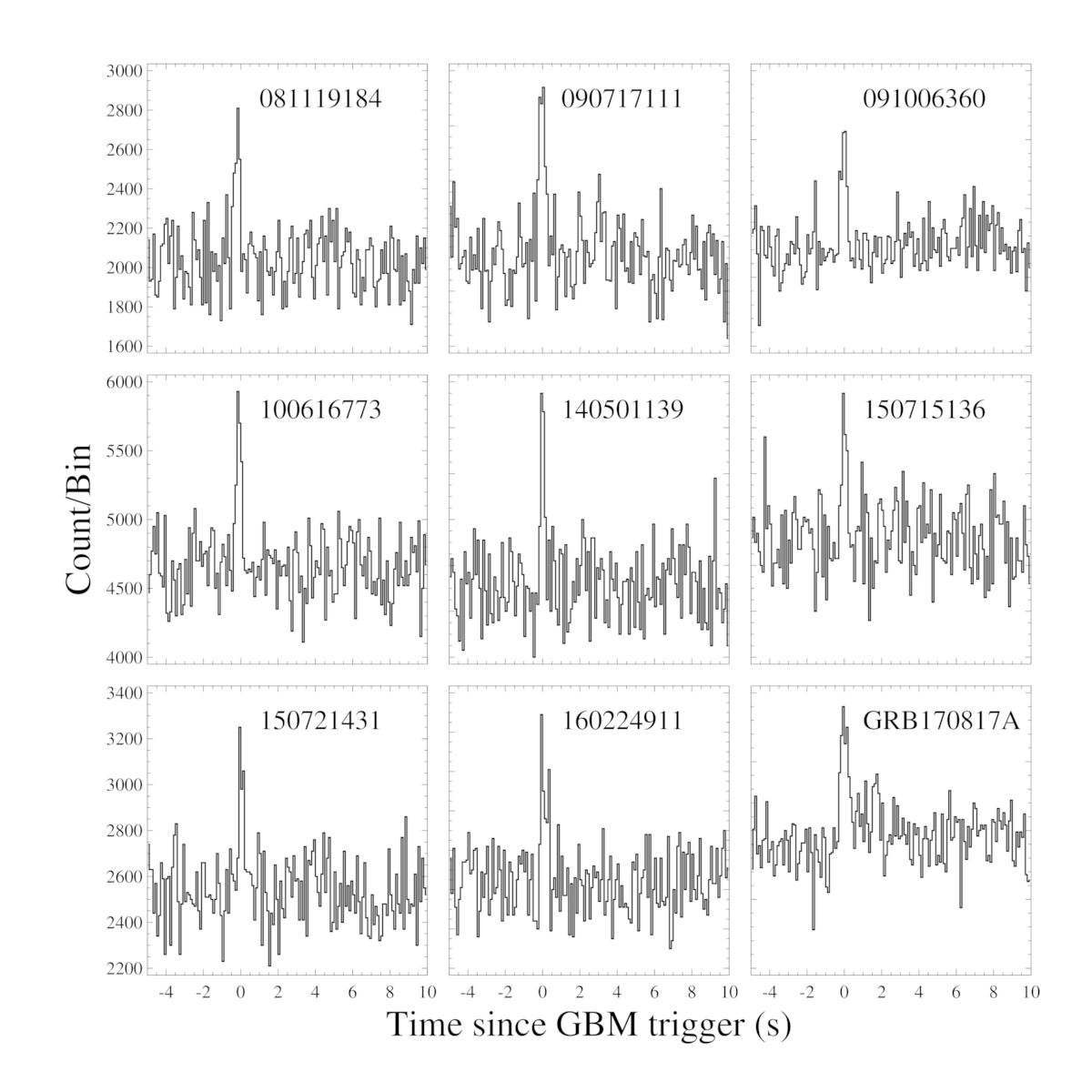} \\

\end{tabular}
\caption{Light curve examples for other GRB 170817A-like events.}
\label{fig:weak-bursts}
\end{figure}

\leftline{\bf Supplementary Note 3. NGC 4993 as a sGRB host}

The host galaxy NGC 4993 of GRB 170817A is an elliptical galaxy in the constellation Hydra. In Supplementary Figure \ref{fig:host} left panel we plot the half light radius $R_{50}$ and stellar mass $M_*$ of NGC 4933 against those of other short GRBs (\cite{li16} and references therein). It is found that NGC 4933 falls in the middle of the distributions and can be regarded as a typical short GRB host. The optical transient SSS17a has a projected distance of 10.6" from the center of NGC 4993. We plot the physical and normalized offset of this event and compare it with other sGRBs (Supplementary Figure \ref{fig:host} right panel). Again, it is consistent with other sGRBs.

\begin{figure}

\begin{tabular}{c}
\includegraphics[keepaspectratio, clip, width=0.7\textwidth]{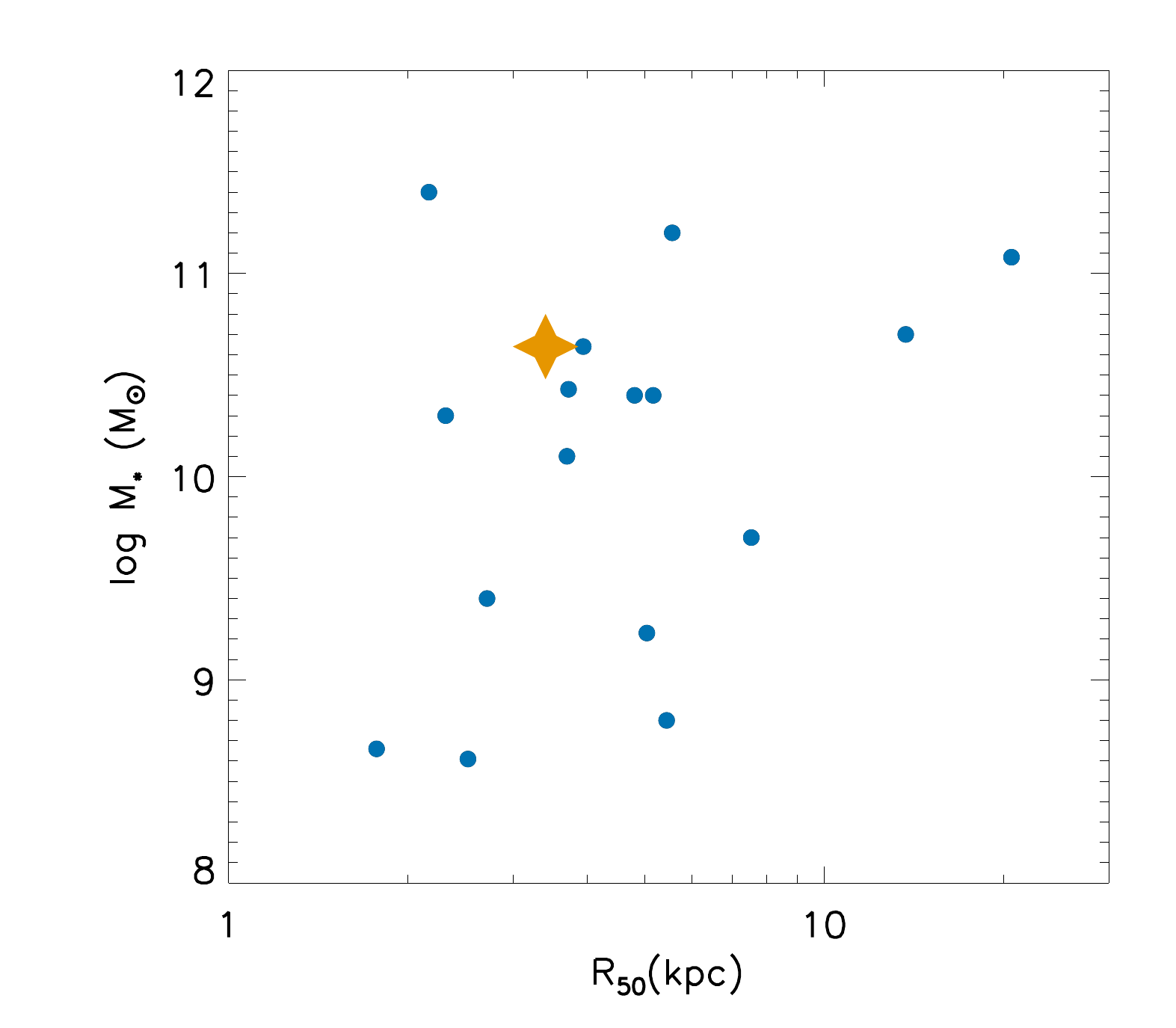} \\
\includegraphics[keepaspectratio, clip, width=0.7\textwidth]{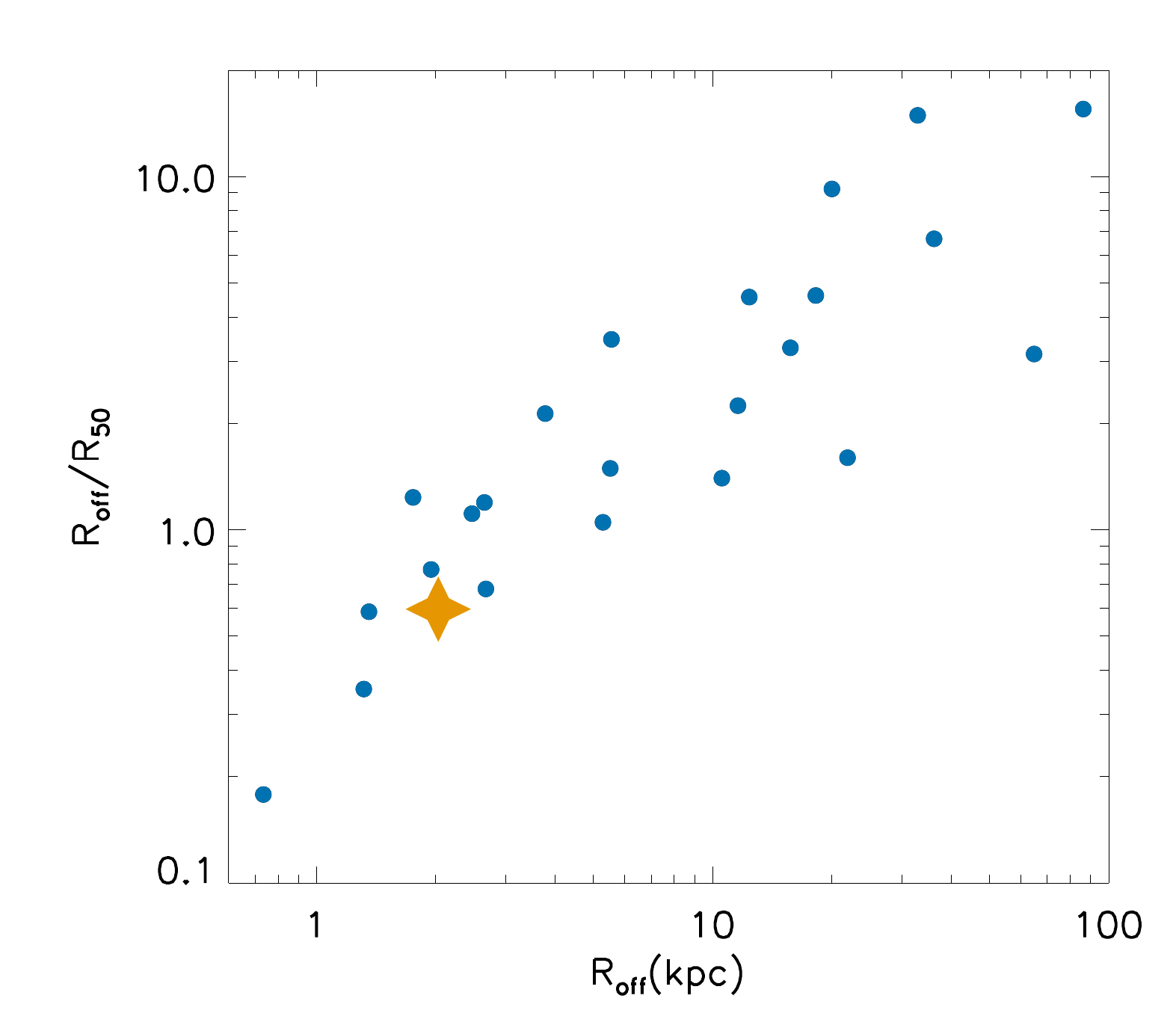} 

\end{tabular}

\caption{{\it Top}: A comparison of the half light radius $R_{50}$ and stellar mass $M_*$ \citep{ogando08} between GRB 170817A and other sGRBs. {\it Bottom}: A comparison of the physical offset $R_{\rm off}$ and normalized offset $R_{\rm off}/R_{50}$ between GRB 170817A and other sGRBs. The dots indicates sGRBs in \cite{li16}. The orange star presents NGC 4993, which falls well into the distributions of the sGRB host galaxy properties. }
\label{fig:host}
\end{figure}

\leftline{\bf Supplementary Note 4. The maximum detectable distance of GRB 170817A}

In order to check at what distance GRB 170817A will become undetectable, we simulate several light curves in 15-350 keV by placing the burst at progressively larger distances from 45 to 80 Mpc. The background level is assumed to be unchanged. The source count rate is assumed
to scale as $D_{\rm L}^{-2}$. A 1-$\sigma$ Poisson noise was added in each simulation. Our simulation suggests that the burst would become hardly detected at $D_{\rm L} \simeq 65$ Mpc (Supplementary Figure \ref{fig:DLmax}). We therefore adopt this value as $D_{\rm L,max}$ to estimate the event rate density of GRB 170817A-like GRBs.

\begin{figure}

\begin{tabular}{c}
\includegraphics[keepaspectratio, clip, width=0.8\textwidth]{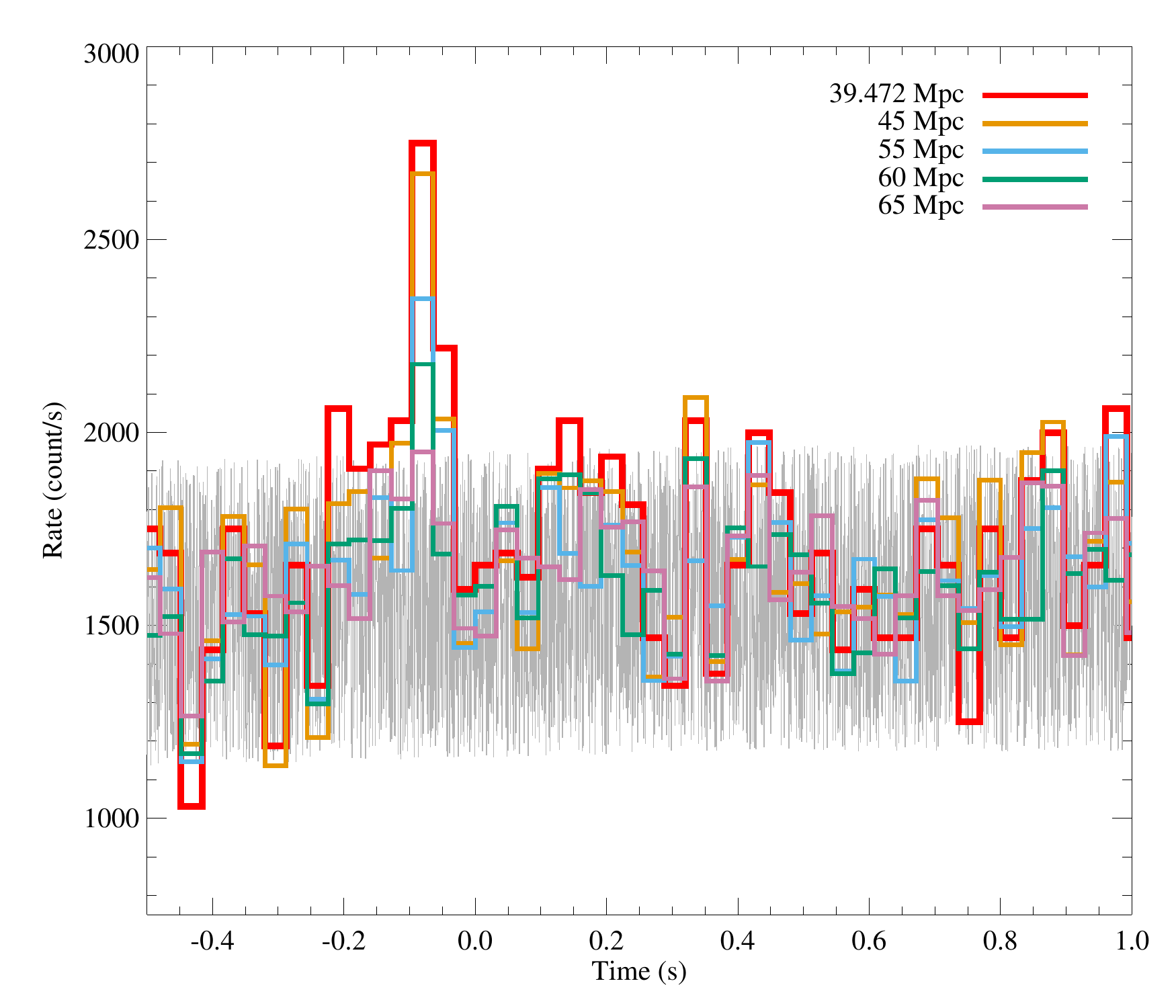} \\

\end{tabular}
\caption{GRB 170817A and simulated bursts by placing it at different distances. A 1-$\sigma$ Poisson noise was added for each simulation. At 65 Mpc, the burst becomes not detectable.}
\label{fig:DLmax}
\end{figure}

\leftline{\bf Supplementary Note 5. Amplitude parameter and possible underlying emission}

The authors of Ref. \cite{lv14} defined an amplitude parameter $f$, which is the ratio between the peak flux and average background flux of a burst. They found that the $f$ parameter can be used to search for disguised short GRBs due to the ``tip-of-iceberg'' effect. Arbitrarily raising the background flux, one can always reduce the duration of a long GRB until its measured duration is shorter than 2 s. The amplitude parameter for such a pseudo GRB was defined as $f_{\rm eff}$ in \cite{lv14}. Comparing the $f$ values of short GRBs and $f_{\rm eff}$ values of long GRBs, \cite{lv14} found that most short GRBs have an $f$ value that is large enough (say, above 2) so that they are genuine. Performing the same analysis to GRB 170817A, we find that its amplitude parameter is relatively small, i.e. $f\sim1.43$.
As shown in Supplementary Figure \ref{fig:f parameter}, this value (red star) is smaller than most short GRBs, and may be confused as a disguised sGRB. The probability ($p$) for it to be a disguised sGRB is $p\sim 0.17$ according to the $p-f$ relation derived by \cite{lv14}. The probability that this burst is an intrinsically short GRB is higher than being a disguised sGRB. Nonetheless, the probability that the intrinsic duration is long is not negligible. One cannot rule out the possibility that there is an underlying, weak, long-duration emission component below the background. In order to search for a possible signal before and after the burst, we perform a detailed spectral analysis in the following time intervals: 
 -2.0 $\sim$ -1.4,
 -1.4 $\sim$ -0.8 ,
 4.0 $\sim$ 13.2 ,
 13.2 $\sim$ 22.4,
 22.4 $\sim$ 31.6,
 31.6 $\sim$ 40.8,
and 40.8 $\sim$ 50.0 seconds. We did not find any significant emission above the background in any of these time intervals and the spectral fitting performed in these intervals simply give overfit/unconstrained parameters.

\begin{figure}
\begin{tabular}{c}
\includegraphics[keepaspectratio, clip, width=0.89\textwidth]{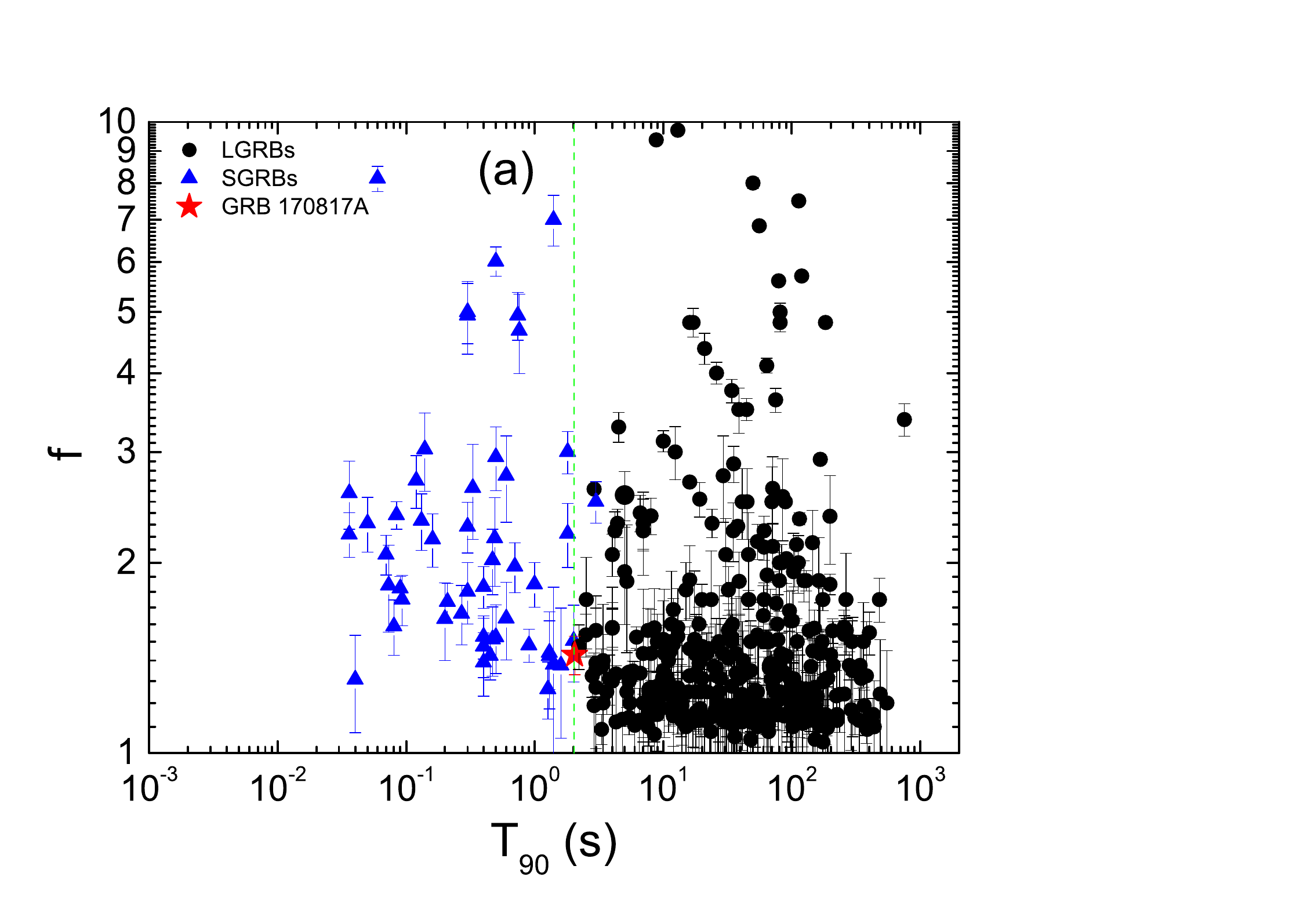}\\
\includegraphics[keepaspectratio, clip, width=0.84\textwidth]{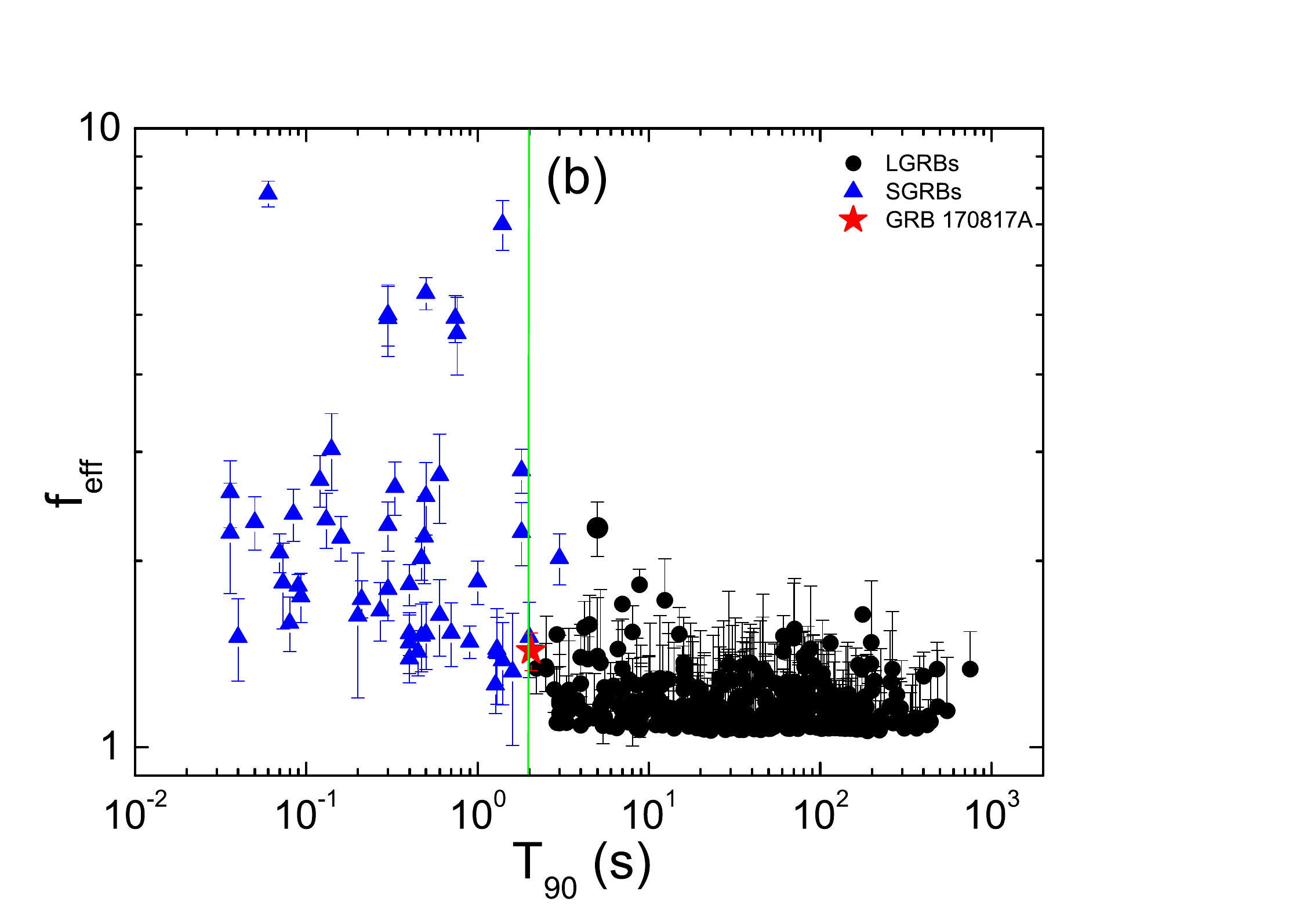}
\end{tabular}
\caption{$T_{90}$ vs. $f$ and $f_{\rm eff}$ diagrams of both long and short GRBs taken from \cite{lv14}.
The red solid star is GRB 170817, and the vertical line is $T_{90}=2$ s. All error bars represent 1-$\sigma$ uncertainties.}
\label{fig:f parameter}
\end{figure}

\leftline{\bf Supplementary Note 6. Off-axis model}

For a uniform jet with a sharp edge viewed outside the jet cone, the observed duration is different from the observed duration of an on-beam observer, given the same central engine activity time scale. The ratio between the two times reads \cite{zhang09}
\begin{equation}
 \frac{t({\rm off-beam})}{t({\rm on-beam})} = \frac{{\cal D}(\theta = 0)}{{\cal D}(\theta=\theta_v-\theta_j)}=\frac{1-\beta \cos(\theta_v-\theta_j)}{1-\beta},
\end{equation}
where $\theta_j$ is the jet opening angle, $\theta_v$ is the viewing angle from the jet axis, ${\cal D} = 1/\Gamma(1-\beta\cos\theta)$ is the Doppler factor, $\beta$ is the dimensionless velocity, $\Gamma$ is the Lorentz factor, and $\theta$ is the angle from the line-of-sight. Since the on-beam Doppler factor can be much larger than the off-beam one, the off-beam duration can be much longer than the on-beam on. { The observed $\sim 2.05$ s duration falls into the distribution of sGRBs. This limits the possible viewing angle of the top-hat model to be at most slightly outside the jet cone. This would suggest that the on-beam luminosity not much brighter than the observed luminosity. This is in conflict with the late time X-ray \cite{troja17} and radio \cite{hallinan17} data, which require significant energetics in the jet axis direction. We therefore conclude that the top-hat jet model viewed outside the jet cone is disfavored by the data.}

{ A more reasonable model invokes a structured jet with angle-dependent isotropic luminosity, with the line of sight piecing through a low-luminosity wing of the jet along which there still exists a relativistic outflow towards the observer. The luminosity structure of the jet can be a simple function (e.g. power law or Gaussian) of the polar angle $\theta$ \cite{zhangmeszaros02,rossi02} or a more complicated function (e.g. two or more components. Within the NS-NS merger scenario, a two-component scenario may be relevant, which invokes a narrow jet beam and a surrounding cocoon material formed to as the jet penetrates through the surrounding dynamical ejecta \cite{lamb17,lazzati17}. In any case, the observed duration in a structured jet scenario is defined either by the central engine activity time scale, or the time scale during which the ejecta radiates, similar to the on-beam case. Invoking a structured jet viewed at a large angle from the jet axis (where the luminosity and bulk Lorentz factor may be both low), one can naturally interprets the duration and low-luminosity of the burst as well as the delayed onset of X-ray and radio emission from the source (e.g. \cite{xiao17}).}

\leftline{\bf Supplementary Note 7. Merger product}

{ The null result of our search for precursor and extended emission before and after the sGRB 170817A is consistent with a BH central engine. Since GW and EM data cannot rule out the possibility of a long-lived neutron star product (e.g. \cite{LIGOproduct}), one may use the available data to constrain the parameter of the underlying NS. Neglecting gravitational wave spindown, the initial dipole spindown luminosity (e.g. \cite{zhangmeszaros01})
$ L_{\rm sd} = 10^{47} \ {\rm erg \ s^{-1}} (B_{p,14}^2 P_{0,-3}^{-4} R_6^6)$
can be constrained by the sGRB luminosity itself.
Here $B_p$ is the surface magnetic field at the polar region, $P_0$ is the initial rotation period, and $R$ is the radius of the pulsar, and the convention $Q=10^n Q_n$ has been adopted in cgs units. Writing $L_\gamma = \eta_\gamma L_{\rm sd}$ where $\eta_\gamma$ is an efficiency parameter, 
in order to satisfy the luminosity constraint, the pulsar needs to have a magnetic field strength $B_p < 7.3\times 10^{13} {\rm G} \ P_{0,-3}^2 \eta_\gamma^{-1/2} $. 
On the other hand, the energy budget of the kilonova and afterglow places a significant limit on the total energy budget of the neutron star. The product has to be either rotating very slowly (which is inconsistent with the expectation of a merger), or has a very low magnetic field together with significant gravitational wave loss more than previously estimated \cite{gao16,piro17} (see a detailed analysis by \cite{yu17}). We therefore suggest that a long-lived neutron star product is less preferred than a BH post-merger product.
}

\leftline{\bf Supplementary Note 8. Delay time between the $\gamma$-ray event and the gravitational wave event}

The merger time derived from the GW signal is T$_{GW}$=12:41:04.4 UTC on 17 August 2017\cite{prl}, which
leads the GRB beginning time by $\Delta t \simeq $ 1.7 s. This delay poses interesting constraints on the GRB emission models.

For a merger that produces a BH, a jet is likely launched promptly after the merger. This is certainly the case for a prompt BH formation,
but would also likely be the case even if there is a short hypermassive neutron star (NS) phase. Even if the system would hold on launching
the jet (say, by $\Delta t_{\rm jet}$), the time scale for a hypermassive NS is typically much shorter than 1.7 s (e.g. 100 ms). 

The BH accretion time scale is short. The fall-back time reads
\begin{equation}
t_{\rm fb} \simeq 2 \left( \frac{R_{\rm out}^3}{G M_\bullet} \right)^{1/2},
\end{equation}
where $R_{\rm out}$ is the outer edge of disk. Usually, we assume the accretion begins at $R_{\rm out} \simeq 2R_{\rm T}$, where $R_{\rm T}$ is the tidal disruption radius with $M_\bullet/R_{\rm T}^3 \sim 4.2\rho_{\rm NS}$. One thus finds,
\begin{equation}
t_{\rm fb} \simeq 2 \left( \frac{2}{G \rho_{\rm NS}} \right)^{1/2} \simeq 5\times 10^{-4} s,
\end{equation}
in which we adopt the typical density for neutron star (NS), i.e., $\rho_{NS} \sim 4\times 10^{14} \rm g \ cm^{-3}$. 

A more relevant timescale is the accretion timescale, which may be estimated as
\begin{equation}
\tau_{\rm acc} \simeq \frac{t_{\rm fb}}{\alpha} \simeq 5 \times 10^{-3} \left(\frac{\alpha}{0.1}\right) \rm s,
\end{equation} 
where $\alpha$ is the viscosity parameter. 
One can see that this time scale is still much shorter than the observed delay.

A longer time scale comes from the propagation of the jet before releasing $\gamma$-rays. The delay time scale depends on the distance
of the emission region from the central engine. In GRB models, both photosphere emission (which corresponds to a small radius
$R_{ph} = L_w \sigma_{\rm T}/(8 \pi \Gamma^3 m_p c^3) \simeq (5.9 \times 10^{10} \ {\rm cm}) \ L_{w,47}^{1/2} \Gamma_1^{-3}$, where $L_w = 10^{47} \ {\rm erg \ s^{-1}} L_{w,47}$ is the wind luminosity, and $\Gamma=10 \Gamma_1$ is the Lorentz factor of the flow) and synchrotron
radiation from an optically thin region ($R_{\rm GRB} \gg R_{ph}$) have been invoked to interpret GRB emission. 

{ Suppose that the GRB emission occurs at radius $R_{\rm GRB}$ with Lorentz factor $\Gamma$, the delay time from the launch of the jet to GRB emission is $t_{\rm prop} \sim R/2\Gamma^2 c$ \cite{zhang16}. 
The total delay time of the onset of the GRB with respect to the GW signal merger time would be 
\begin{equation}
\Delta t \sim (t_{\rm prop} +\tau_{\rm acc} + \Delta t_{\rm jet} ) (1+z) \simeq (\tau_{\rm prop} + \Delta t_{\rm jet} ) (1+z). 
\end{equation}
An intriguing fact is that the GRB duration $T_{90} \sim 2$ s, which is similar to the delay time scale $\Delta t \sim 1.7$ s. If one does not introduce an ad hoc $\Delta t_{\rm jet} \sim 1$ s by hand (e.g. as invoked in the cocoon breakout model or the photosphere model \cite{cocoon}), one natural interpretation is that $\Delta t \sim t_{\rm prop} \sim T_{90} \sim R_{\rm GRB}/\Gamma^2 c$. Adopting $\Delta t = 1.7$ s from the data, the emission radius may be estimated as
\begin{equation}
R_{\rm GRB} \sim \Gamma^2 c t =5 \times 10^{14} {\rm cm} \left(\frac{\Gamma}{100}\right)^2 \left(\frac{\Delta t}{1.7 {\rm s}} \right)= 5 \times 10^{12} {\rm cm} \left(\frac{\Gamma}{10}\right)^2 \left(\frac{\Delta t}{1.7 {\rm s}} \right),
\end{equation} 
which is usually much greater than the photosphere radius $R_{ph}$. The photosphere emission may give such a delay if $\Gamma <5$. However, the temperature of the photosphere emission would be too low to explain the high value of $E_p \sim 158$ keV. If the emission is not from the photosphere, then the photosphere emission has to be suppressed via magnetization (e.g. \cite{zhangpeer09}). One needs a Poynting flux dominated flow, advected to a large radius before magnetic dissipation happens (e.g. \cite{zhangyan11}). As a magnetic bubble penetrates throgh the surrounding cocoon, a jet structure naturally develops. An observer at a viewing angle $\theta_v \sim 28$ deg would observe a low-luminosity GRB with delay time scale comparable to the duration itself regardless of the unknown values of the Lorentz factor $\Gamma$ and emission radius $R$.}

\leftline{\bf Supplementary Note 9. Predicted afterglow properties}

The interaction between the jet and its ambient medium could generate a strong external shock, where particles are believed to be accelerated, giving rise to broad-band afterglow emission \cite{gao13review}. According to standard afterglow models, the lightcurve for a given observed frequency (e.g. optical frequency $\nu_{\rm obs}$) could be calculated as
\begin{equation}
F_{t,\nu_{\rm obs}} = f(t,\nu_{\rm obs}; z, p, n, \epsilon_e, \epsilon_B, E_k, \Gamma_{0}),
\label{eq:para}
\end{equation}
where $E_k$ is the isotropic kinetic energy of the jet, $\Gamma_0$ is the initial Lorentz factor of the jet and n is the interstellar medium (ISM) particle number density. $\epsilon_e$ and $\epsilon_B$ are the electron and magnetic energy fraction parameters, and $p$ is the electron spectral index. 

Based on the total emission energy of the prompt emission and assume a factor of $20\%$ for the $\gamma$-ray emission efficiency, the kinetic energy of the jet $E_k$ can be estimated as $1.83\times10^{47}~{\rm erg}$. For binary neutron star mergers, a low value for ambient medium density is usually expected, since they tend to have a large offset relative to the center of its host galaxy. Here we adopt the ambient medium density $n$ as $10^{-3}~{\rm cm^{-3}}$. For a structured jet viewed from a large angle, the initial Lorentz factor may be low, so we adopt $\Gamma_0 = 20$. For other parameters, we adopt their commonly used values in GRB afterglow modeling, i.e., $\epsilon_e=0.1$, and $p=2.3$ \cite[][for a review]{kumarzhang15}. The distribution of the $\epsilon_B$ value is wide, from $\epsilon_B=0.01$ to $\epsilon_B < 10^{-5}$ \citep{santana14,wang15,beniamini15,2015ApJ...806...15Z,2015ApJ...799....3R}.

We use $\epsilon_B=0.01$ to calculate the most optimistic case of afterglow emission. The peak flux of the X-ray 
light curve emerges around 2000s, at the level of $10^{-14}~{\rm erg/cm^2/s}$. Around 1 day, the X-ray flux will decay to the level of $10^{-15}~{\rm erg/cm^2/s}$, under the detection limit of Swift/XRT. This is consistent with the non-detection result by the {\em Swift} team \cite{Evans17}. For optical band, the peak of the light curve appears around 2000 s, and the peak flux is $1~\mu Jy$ (AB magnitude is 24). Around 1 day, the optical flux decays to the level of $0.01~\mu Jy$ (AB magnitude is 28). This is much lower than the observed flux. This suggests that the optical transient detected by multiple groups originates from the emission of a kilonova \citep{li98,kulkarni05,metzger10,yu13}. { The late emergence of the X-ray and radio emission \cite{troja17,hallinan17} are consistent with emission from the near-axis powerful jet being decelerated by the ambient medium \cite{xiao17}.}

\leftline{\bf Supplementary Note 10. Search for GRB 170817A-like events}

Since it is possible that there are other GRB 170817A-like events in the GBM faint sGRB sample, we attempted to search for these events using the galaxy data. We choose all the faint bursts listed in Supplementary Figure \ref{fig:weak-bursts}, and look for NGC 4993-like galaxy (with luminosity $3.6 \times 10^{43} \ {\rm erg \ s^{-1}}$) below 80 Mpc within the error boxes of the sGRBs. Supplementary Figure \ref{fig:search} shows the galaxies (red dots) with luminosity $>10^{43} \ {\rm erg \ s^{-1}}$ (left) and $>3\times 10^{43} \ {\rm erg \ s^{-1}}$ (right) compared with sGRB error circles. It is clearly seen that the error circles are too large and typically enclose many galaxies. So identifying GRB 170817A-like events is difficult without gravitational wave detections.

\begin{figure}
\begin{tabular}{c}
\includegraphics[keepaspectratio, clip, width=0.85\textwidth]{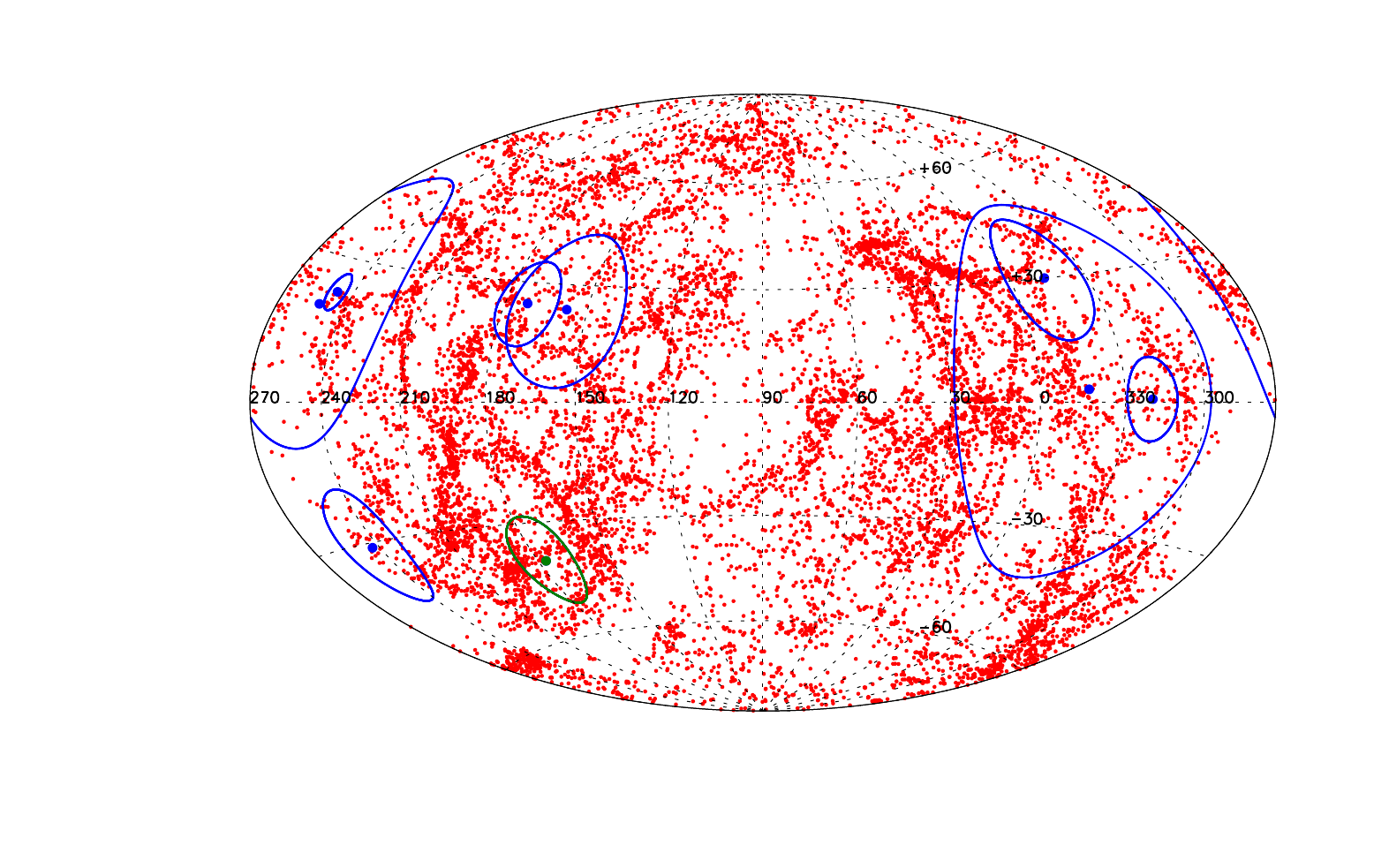} \\
\includegraphics[keepaspectratio, clip, width=0.85\textwidth]{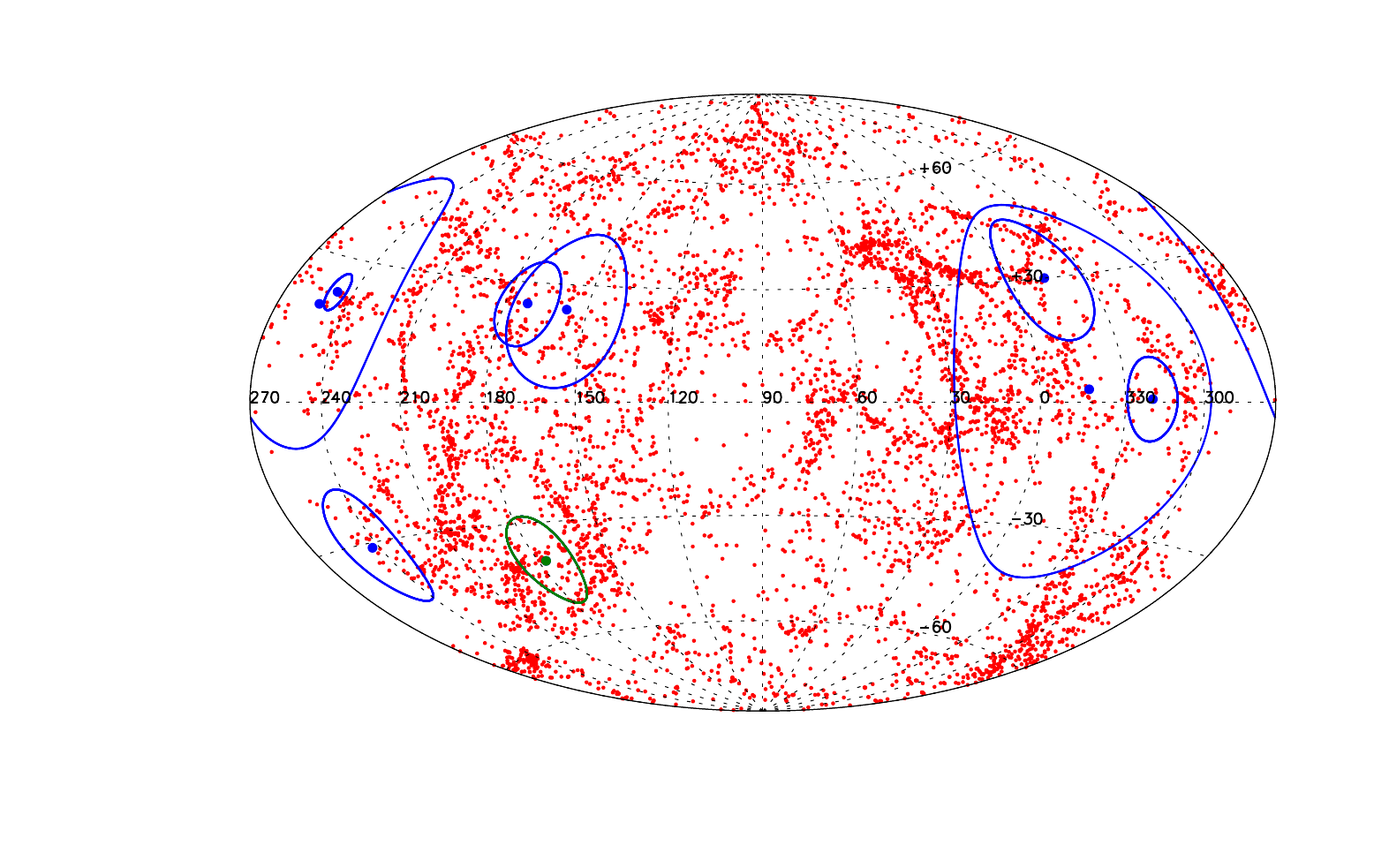} 

\end{tabular}
\caption{The position error circles of GRB 170817A-like sGRBs in Figure \ref{fig:weak-bursts} compared with the sky map of galaxies below 80 Mpc with two different luminosity threshold: $>10^{43} \ {\rm erg \ s^{-1}}$ (top) and $>3\times 10^{43} \ {\rm erg \ s^{-1}}$ (bottom). The green color indicates GRB 170817A.}
\label{fig:search}
\end{figure}

\begin{figure}
\begin{tabular}{c}
\includegraphics[keepaspectratio, clip, width=0.65\textwidth]{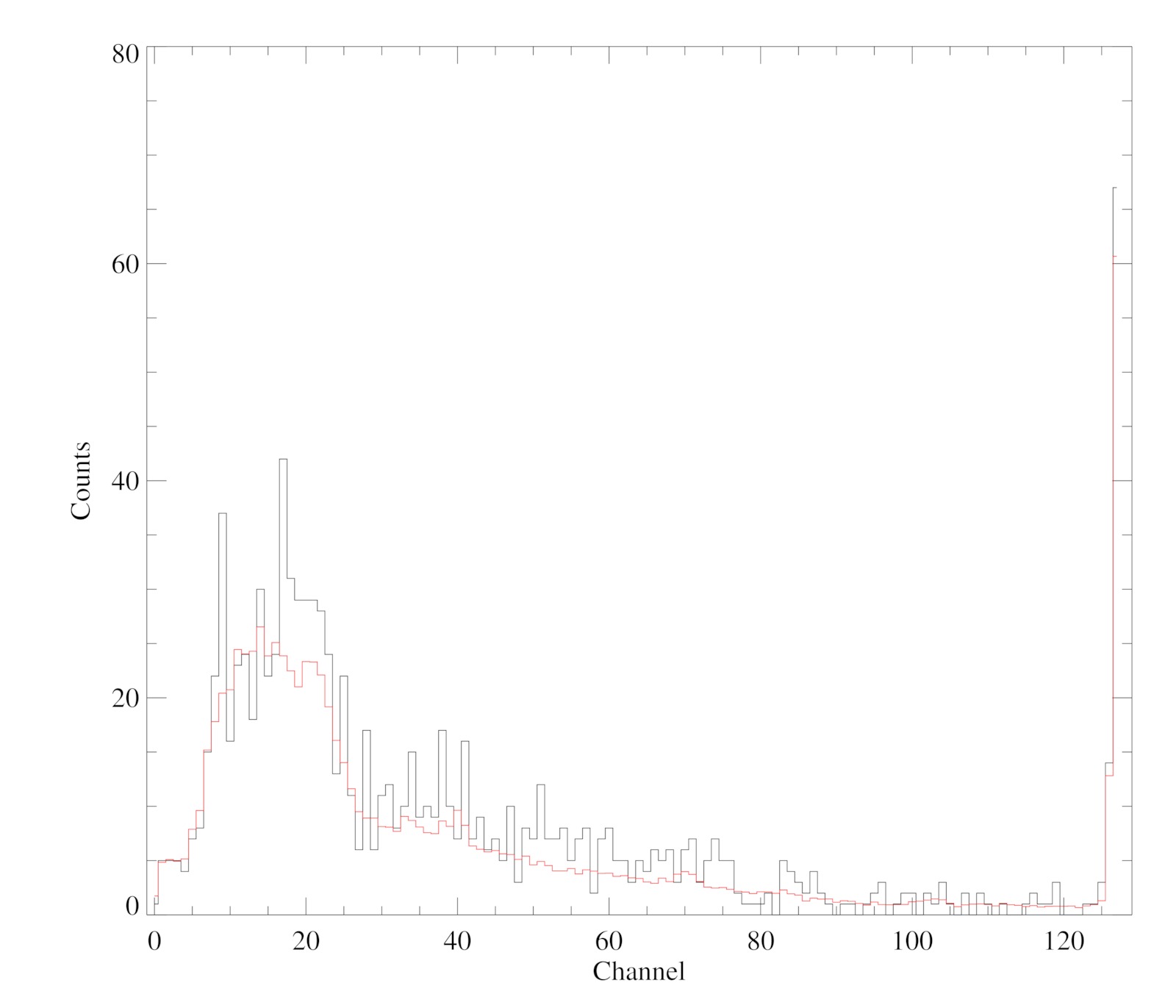} \\
\includegraphics[keepaspectratio, clip, width=0.65\textwidth]{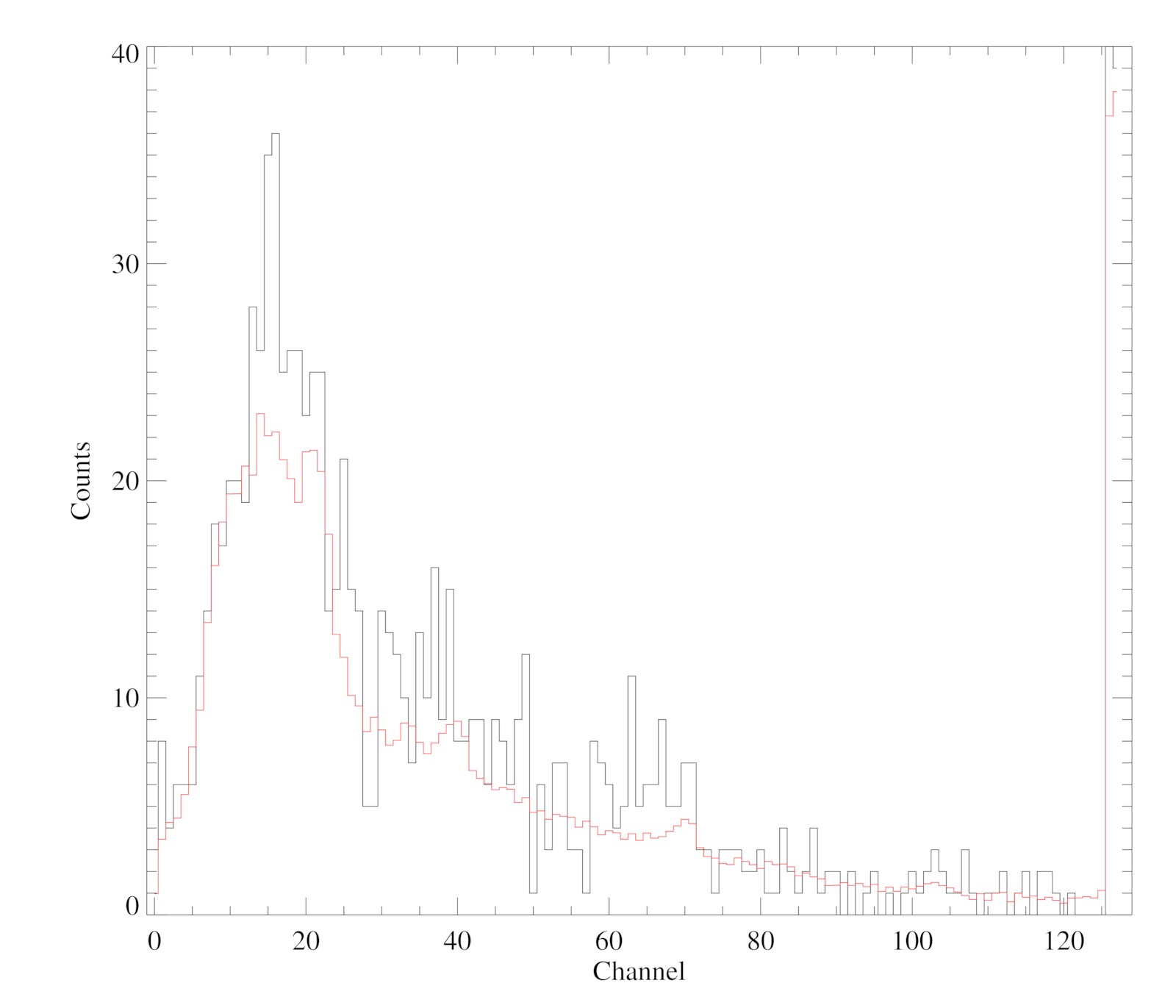} 

\end{tabular}
\caption{Total counts (black) vs background counts (red) in each channel of the GBB NaI detectors in interval -0.26 to 0.57 s. {Top:} detector n1. The total counts over all the channels is 1048 , the total background counts over all the channels is 862. {bottom:} detector n2. The total counts over all the channels is 999 , the total background counts over all the channels is 822. }
\label{fig:source_bak_cts}
\end{figure}

\clearpage

\vspace{\baselineskip}
\textbf{Supplementary References}

\makeatletter
\def\@biblabel#1{\@ifnotempty{#1}{#1.}}

\end{document}